\documentclass{jfm}
\usepackage{graphicx}
\usepackage{epstopdf, epsfig}
\usepackage{subcaption}
\usepackage[section]{placeins}
\usepackage[export]{adjustbox}

\usepackage{gensymb}
\usepackage{amsmath}
\usepackage{dirtytalk}
\usepackage[section]{placeins}
\usepackage{needspace}
\usepackage{color}

\usepackage{siunitx}

\shorttitle{Micro-droplet nucleation through solvent exchange in a turbulent buoyant jet}
\shortauthor{You-An Lee, Chao Sun, Sander G. Huisman, Detlef Lohse}

\title{Micro-droplet nucleation through solvent exchange in a turbulent buoyant jet}

\author{You-An Lee\aff{1}
 , Chao Sun\aff{1,2}
 , Sander. G. Huisman\aff{1},
 \and Detlef Lohse\aff{1,3}}

\affiliation{\aff{1}Physics of Fluids Group and Max-Planck, Center for Complex Fluid Dynamics, Faculty of Science and Technology, J.M. Burgers Center for Fluid Dynamics, University of Twente, 7500 AE Enschede, Netherlands

\aff{2}Center for Combustion Energy, Key Laboratory for Thermal Science and Power Engineering of Ministry of Education, Department of Energy and Power Engineering, Tsinghua University,
100084 Beijing, China

\aff{3}Max Planck Institute for Dynamics and Self-Organization, Am Fa\ss berg 17, 37077 G\"{o}ttingen,Germany
}

\begin{document}

\maketitle

\begin{abstract}
Solvent exchange is a process involving mixing between a good solvent with dissolved solute and a poor solvent. The process creates local oversaturation which causes the nucleation of minute solute droplets. Such ternary systems on a macro-scale have remained unexplored in the turbulent regime. We experimentally study the solvent exchange process by injecting mixtures of ethanol and trans-anethole into water, forming a turbulent buoyant jet in upward direction. Locally, turbulent mixing causes oversaturation of the trans-anethole following turbulent entrainment. We optically measure the concentration of the nucleated droplets using a light attenuation technique and find that the radial concentration profile has a sub-Gaussian kurtosis. In contrast to the entrainment-based models, the spatial evolution of the oversaturation reveals continuous droplet nucleation downstream and radially across the jet, which we attribute to the limited mixing capacity of the jet. Though we are far from a full quantitative understanding, this work extends the knowledge on solvent exchange into the turbulent regime, and brings in a novel type of flow, broadening the scope of multicomponent, multiphase turbulent jets with phase transition. 
\\
\end{abstract}

\section{Introduction}
Solvent exchange can be employed to extract specific components from mixtures. The process finds various industrial applications, such as liquid-liquid extraction \citep{Lohse2016}, drug delivery \citep{Lepeltier2014}, resource recycling \citep{Zimmermann2014}, and oil recovery \citep{Sun2017}. Such solvent exchange processes consist of a ternary liquid system, where mixing between a good solvent and a poor one takes place, leading to oversaturation of the solute originally dissolved in the good solvent. This in turn triggers nucleation of the solute droplets. Such droplet nucleation is called spontaneous emulsification, or is known by its popular name `ouzo effect'. In this process metastable droplets are formed in the liquid, creating a milky appearance due to Mie scattering for the case of watered-down ouzo beverage. In recent decades, numerous efforts have been made to understand such a complex multicomponent and multiphase flow, see the review by \citet{Lohse2020}. \citet{Vitale2003} and \citet{Sitnikova2005} were among the earliest studies of the ouzo effect, using divinylbenzene and trans-anethole as the solute, respectively. They both located the position of the metastable region in the ternary phase diagram, and experimentally revealed the diffusive growth process of the droplets and their size distribution. 

To quantitatively study the diffusion dynamics of the multicomponent droplets,  \citet{Su2013} extended the classic work of \citet{Epstein1950} on the dissolution and growth of gas bubble in surrounding liquid to the case of a  droplet. \citet{Chu2016} further analytically investigated the droplet growth and dissolution of the multicomponent droplets, broadening the theoretical framework of droplet diffusion dynamics, making it more applicable to the real-world applications, such as liquid-liquid extraction. Corresponding molecular dynamics simulation was done by \citet{Maheshwari2017}. \citet{Tan2019} studied the micro-droplet nucleation of the water-ethanol drop in host anise oil. They combined the multicomponent diffusion model, thermodynamic equilibrium, and a scaling analysis of Marangoni and buoyancy forces. With this they could successfully predict the spontaneous emulsification by the convection-enhanced diffusion process. 

While the aforementioned cases are homogeneous nucleation in the bulk, the ouzo effect has also been widely investigated on the surface of microfluidic channels, so-called surface droplet formation via heterogeneous nucleation.  \citet{Hajian2015} studied the droplet nucleation and its radial migration in a micro-channel, measuring the concentration field formed by the solvent exchange and the following diffusive mass transfer. For controlled solvent exchange in a microfluidic setups, \citet{Zhang2015} found that the volume of the nucleated droplets scales as $Pe^{3/4}$, with Peclet number $Pe = \frac{Q}{wD}$, where $Q$ is the volume flow rate, $w$ the channel width, $D$ the diffusion coefficient. Following this finding, the universality of the phenomena has been demonstrated, including the effect of the solution composition \citep{Lu2015, Lu2016}, confining the flow geometry \citep{Lu2017, Zeng2019}, and formation \citep{Li2018} and growth \citep{Dyett2018} of the surface nanodroplets. For related work, \citet{Li2021} implemented a Hele-Shaw like thin cell with a porous section to better collect the nucleated droplets. For this geometry,  they formulated a theoretical model based on the ternary diagram and the diffusion equation, and found and explained the scaling of the volume of the nucleated droplets in the oversaturation front to be proportional to $Pe^{1/2}$.

While solvent exchange is thus reasonably and in part even quantitatively understood, the scope of quantitative studies has up to now been limited to small scales and the laminar flow regime. For industrial-scale applications such as liquid-liquid extraction, however, processes on larger scales are generally less controlled than those in microfluidic devices due to turbulence. It is therefore highly desirable to unveil the diffusion dynamics of the ternary liquid system in turbulent environments, where intense mixing and nucleation appear. The turbulent jet is one of the most fundamental and well-studied form of turbulent shear flow (see e.g. \citet{Pope2000}), which makes it an ideal candidate to bridge the field of solvent exchange and turbulence.

Built upon the classic single phase studies for turbulent jets and plumes \citep{List1982, Morton1956, Fischer1979, Turner1986, Pope2000}, the case of multiphase flow has drawn extensive interest due to its common occurrence in environmental fluid mechanics \citep{Woods2010} and industrial applications such as combustion \citep{Raman2016} and aerosol formation \citep{Lesniewski1998, Neuber2017}. The Deepwater Horizon Oil Spill further triggered a series of investigations into such types of multiphase flow \citep{Mcnutt2012,Reddy2012,Ryerson2012}. With turbulent entrainment, bubbles or droplets can grow and dissolve in a jet, inducing significant buoyancy variation in the flow. \citet{Chu2019} analytically identified key parameters for the multiphase plume dynamics, including the droplet dissolution rate, the density change rate due to dissolution, the droplet velocity, and the plume velocity. 

In addition to the dissolution, considerations for chemical reaction have advanced the line of research further to tackle real-world complex problems. There are two major perspectives of research on reactive jets and plumes. On the one hand, the ambient fluid reacts with the dispersed phase in the jet or plume. This category serves as an extension of the works on bubbles or droplets dissolution, focusing on the buoyancy variation accompanying the reactions, see \citet{CARDOSO2010} and \citet{Domingos2013}. These authors also discussed the competition between stratification and reaction as the dominant source of buoyancy variation. 

On the other hand, the ambient fluid reacts with the carrier phase of the jet, or simply with the single phase injected fluid. Studies on such type of flow shed new light on the interplay among entrainment, mixing, and reactions as the flow develops, enabling the disentanglement of these effects by comparing the corresponding length and time scales. \citet{Ulpre2013} injected a single phase acidic jet into an alkaline environment, reporting a model combining chemistry and the fluid dynamics of turbulent plumes to predict the neutralization distance. \citet{Domingos2015} studied single phase reacting thermals, distinguishing fast and slow reactions. Their experimental findings show that the reactions only occurred preferentially in part of the thermals for the instantaneous case, while for the slow reactions the thermals were homogeneous in composition. \citet{Mingotti2019b} experimentally compared the slow and instantaneous reaction in turbulent plumes, discovering preferential reaction for the instantaneous case as well, which appeared on the edge of the eddies. They formulated a theoretical prediction for the concentration evolution in reacting plumes using scale analysis of entrainment, mixing, and reactions. Solvent exchange in our case is fundamentally different from a chemical reaction in terms of driving mechanism. The product of solvent exchange, namely the nucleated droplets, is driven by the degree of oversaturation, while the products of chemical reactions result from chemical kinetics. Also, speaking of chemical reaction and mixing, \citet{Guilbert2021a} experimentally studied the effect of reaction on the concentration distribution of the product in a randomly stirred mixture, distinguishing between diffusion-controlled and reaction-controlled regimes \citep{Guilbert2021b}, and showing a compressed distribution profile for reactive mixtures \citep{Guilbert2021a}. In spite of such difference, the cases of instantaneous reactions in the aforementioned studies are highly insightful for our attempt to understand the solvent exchange process in turbulent buoyant jet, not only because the timescale to generate the product are both extremely small, but also because the turbulent entrainment and mixing might play key roles in both processes. 

The present experimental study aims to investigate the solvent exchange process in a multiphase turbulent buoyant jet in the upward direction, bridging the gap between the ouzo effect and turbulent shear flow. We will employ trans-anethole and refer to it as $oil$ throughout the paper as it is the oil phase in the ternary liquids system. We will use a vertical ethanol-oil jet being injected in a quiescent water bath, creating an 'ouzo' mixture close to the needle. We utilize a novel method to construct the averaged oversaturation (concentration) field for the ouzo solution using titration, light attenuation technique, and minimization using an axisymmetric discretization. We compare the ouzo case with the nucleation-free case of injecting only dyed ethanol, revealing the distinct features of the ouzo case in the radial concentration profile, the centerline evolution, and the oversaturation flow rate. Adapting the mixing-limited arguments for the prolonged depletion of the reacting plume of \citet{Mingotti2019b}, the extended lifetime of the droplet-laden respiratory puff of \citet{Chong2021} and of the evaporating spray of \citet{Rivas2016} and \citet{Villermaux2017}, we provide a qualitative explanation for the continuous nucleation downstream and the wide-spread location of the micro-droplet nucleation across the jet.  

The paper is structured as follow: In \S2 we detail the experimental setup and the methods to measure the velocity field and the concentration field, and to resolve the nucleated droplets. The concentration fields can be reconstructed by combining the information from two cameras, as presented in \S3. The results extracted from the concentration fields are presented in \S4, with supporting discussion and qualitative analysis in \S5. Conclusions and future prospects are presented in \S6.

\section{Experimental method}
\subsection{Setup}
\begin{figure}
    \centering
    \includegraphics[scale=0.7]{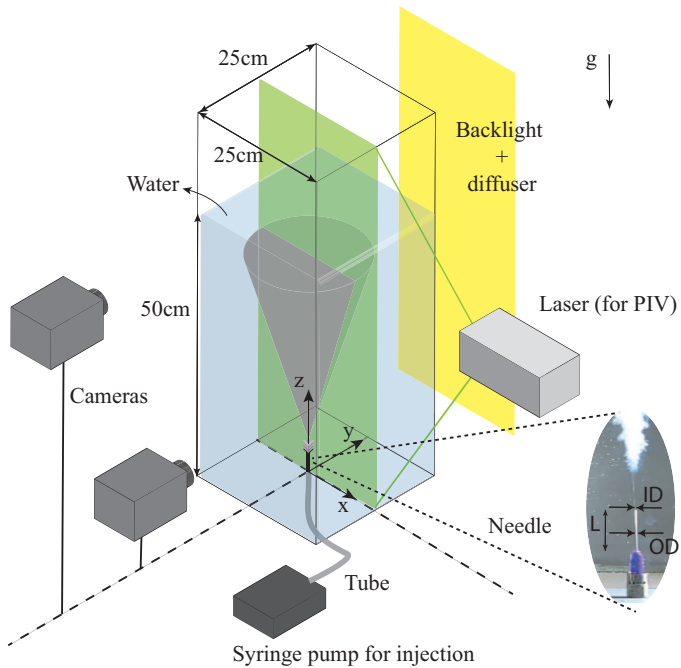}
    \caption{Experimental setup. The inset shows the photo of the needle for injection with inner diameter (ID), outer diameter (OD), and length (L). Note that the laser was only activated for PIV.}
    \label{img:setup}
\end{figure}

We have conducted several experiments to study the solvent exchange process in a turbulent buoyant jet (ouzo jet). The jet experiments were operated in a glass tank with dimensions \SI{25}{\cm} $\times$  \SI{25}{\cm} $\times$  \SI{50}{\cm} (W $\times$ L $\times$ H)  as in Fig. \ref{img:setup}. The tank was filled with decalcified water before each run of experiments. The injected fluid consisted of mixtures of ethanol (Boom, 100\%) and trans-anethole (Sigma Aldrich, $\geq$99\%) with varying weight ratio, and the fluid was injected by a Harvard 2000 syringe pump at selected flow rates, forming an upwards turbulent buoyant jet through a round needle with inner diameter  \SI{0.51}{\mm}, outer diameter \SI{0.82}{\mm}, and length \SI{12.7}{\mm}. Reference cases were also conducted by injection of dyed ethanol, offering comparison between the common passive scalar jet and the ouzo jet. The water and the injected liquid were kept at room temperature ($\SI{20}{\celsius}\pm \SI{1}{\kelvin}$) for all the experiments to minimize the effect of the temperature-dependence of the solubility in the ternary liquids system. The experimental conditions are listed in Table \ref{tbl:condition}. 

To measure the local concentration we use a light attenuation technique. For this we use a LED back lighting with diffuser and recorded using  two Photron FASTCAM Mini AX200 high speed cameras with 1024 $\times$ 1024 pixels resolution at 50 fps, one from afar with a Sigma \SI{50}{\mm} objective to capture the entire jet evolution, and the other with a Zeiss \SI{100}{\mm} objective placed closer to the tank and zoomed in to the region right above the injection needle. The injection and the recording lasted for 40--50 s, providing 2000--2500 frames for each experiment. To obtain the velocity field of the turbulent buoyant jets, we relied on particle image velocimetry (PIV) with laser activation and using different cameras, which will be detailed in \S2.5.
\begin{table}
\begin{center}
\def\arraystretch{1.5}
\setlength{\tabcolsep}{10pt}
\begin{tabular}{r r r r r r r} 
No. & $w_e/w_o$ & $C_{\text{dye}}$ (ppm) & \vtop{\hbox{\strut $Q$ $(\text{m}^{\text{3}}/\text{s})$}\hbox{\strut $\times 10^{-7}$}} & \vtop{\hbox{\strut $L_m$ (m)}\hbox{\strut $\times 10^{-2}$}} & $u_0$ (m/s) & $Re_0$ \\ 
1 & 100 & - & 3.33 & 2.48 & 1.63 & 555  \\ 
2 & 100 & - & 8.33 & 6.19 & 4.08 & 1387 \\
3 & 33 & -  & 3.33 & 2.48 & 1.63 & 555\\
4 & 33 & -  & 8.33 & 6.19 & 4.08 & 1387\\
5 & - & 4000  & 3.33 & 2.48 & 1.63 & 555\\
6 & - & 4000  & 8.33 & 6.19 & 4.08 & 1387\\ 
\end{tabular}
\end{center}
\caption{Experimental conditions. Each experiment was repeated three times to confirm the repeatability of the data. Experiments 1--4 correspond to the ouzo jets with empty $C_{\text{dye}}$ column, and experiments 5--6 correspond to the dyed ethanol jets with empty $w_e/w_o$ column. $w_e/w_o$ denotes the weight ratio between ethanol and the oil in the injected fluid. $Q$ is the volume flow rate regulated by the syringe pump, leading to different characteristic length $L_m$,  initial velocity $u_0$, and initial Reynolds number $Re_0=Qd/(\frac{1}{4}\pi d^2 \nu)$, where $d=0.51$ mm is the inner diameter of the needle. Here, $L_m = M^{3/4}/B^{1/2}$, where $M=Q^2/(\frac{1}{4}\pi d^2)$ is the initial momentum flux, and $B=Qg(\rho_{\text{jet}}-\rho_{\text{amb}})/\rho_{\text{amb}}$ is the initial buoyancy flux. $B=1.635 \times 10^{-7}$ $\text{m}^{\text{4}}/\text{s}^{\text{3}}$ is constant throughout our experiments as the density of the oil is negligible, i.e., we consider the jet density $\rho_{\text{jet}}$ to be solely determined by the ethanol. The viscosity of the oil was also neglected, leading to a constant viscosity $\nu = \nu_{\text{e}} = 1.5 \times 10^{-6}$ $\text{m}^{\text{2}}/\text{s}$, which is used to calculate $Re_0$.}
\label{tbl:condition}
\end{table}
\subsection{Titration and oversaturation}

\begin{figure}
\centering
\begin{subfigure}[t]{0.5\textwidth}
\caption{}
\includegraphics[scale=0.8]{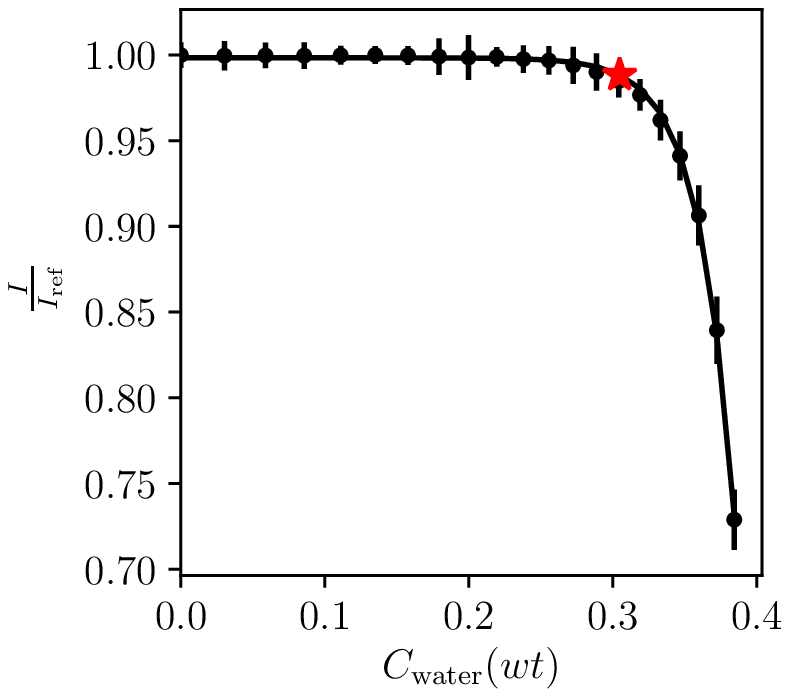}
\label{img:titration}
\end{subfigure}\hfill%
\begin{subfigure}[t]{0.5\textwidth}
\caption{}
\includegraphics[scale=0.4]{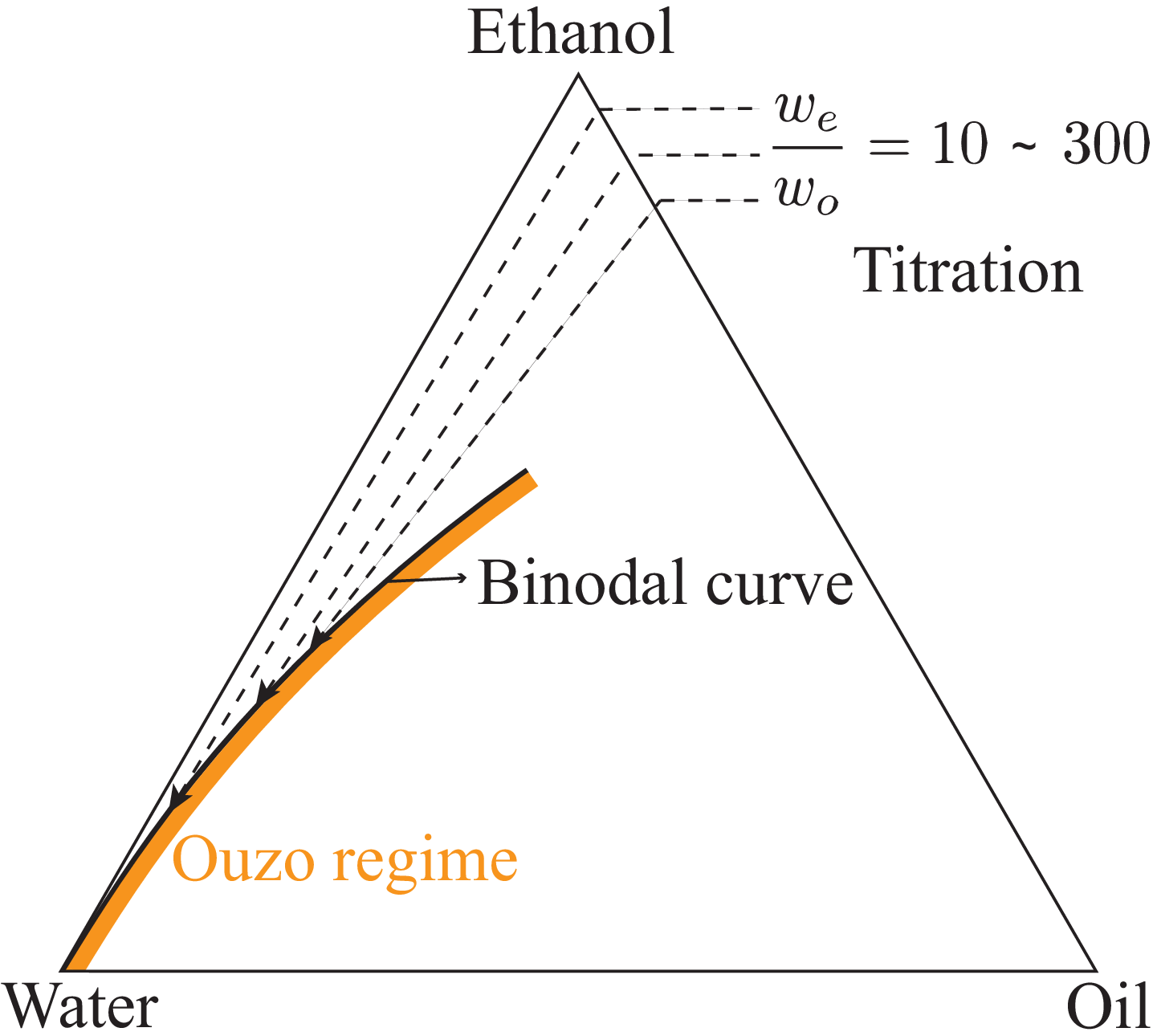}
\label{img:ternary}
\end{subfigure}\vspace{-3.5mm} 
\begin{subfigure}[t]{0.5\textwidth}
\caption{}
\includegraphics[scale=0.8]{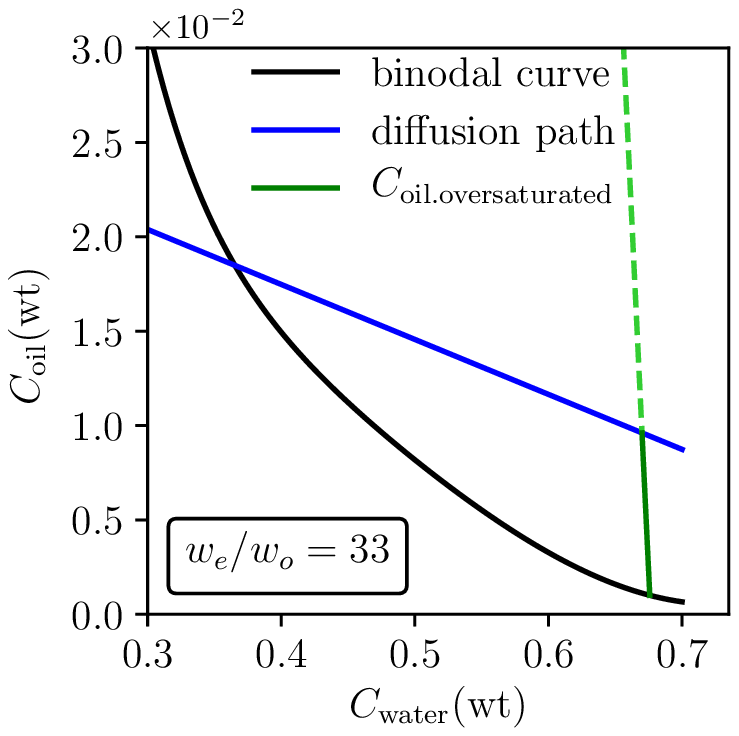}
\label{img:binodal}
\end{subfigure}%
\begin{subfigure}[t]{0.5\textwidth}
\caption{}
\includegraphics[scale=0.8]{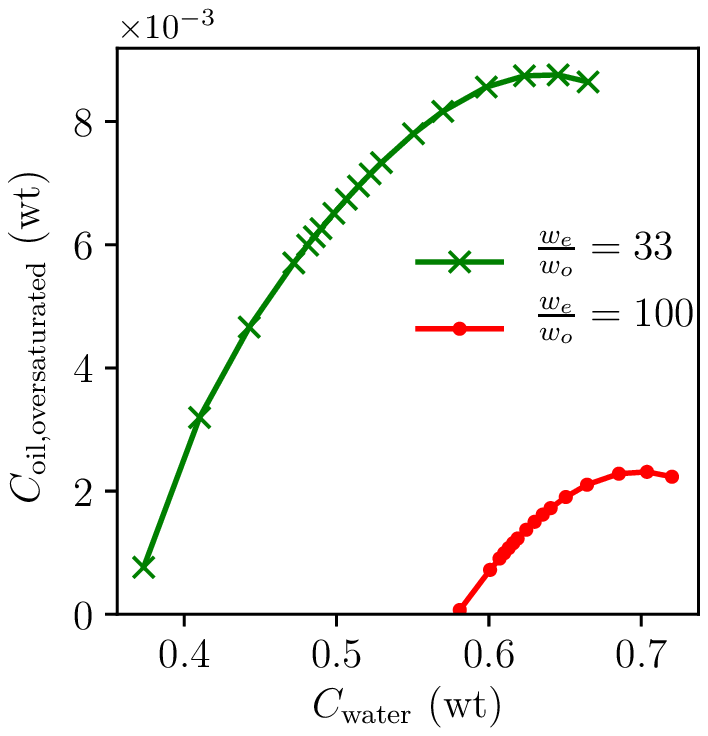}
\label{img:oversat}
\end{subfigure}%

\caption{(a) Camera assisted titration to determine the saturation point for a specific $w_e/w_o$. The black data points are obtained by averaging the light intensity of the entire domain, and the error bar denotes the standard deviation of the light intensity within the domain. The star marks water concentration ($C_{\text{water}}$) causing 1\% drop of the light intensity. (b) A simplified ternary diagram to illustrate the titration process and the constructed binodal curve.} (c) show the binodal curve and the theoretical diffusion path for the case $w_e/w_o=33$. (d) The relation between the water concentration and oil oversaturation can then be built from (c) for the following calibration.
\label{img:saturation}
\end{figure}

To study the nucleation and the solvent exchange process quantitatively, knowledge of oversaturation of the nucleated component is required \citep{Li2018, Li2021}. This means that the first step is to obtain the binodal curve, namely the saturation curve by titration. Normally, the titration results of such ternary system are determined visually by checking the change of appearance. However, this is extremely difficult since we operated in the regime with extremely small amounts of oil. We therefore opted for cameras for accurate determination of the point of saturation. The camera-assisted titration was conducted by step-wise adding water into the oil-ethanol solution, followed by one minute thorough mixing, and then idle time for around 15 minutes to ensure the temperature stabilizes to $\SI{20}{\celsius}\pm \SI{1}{\kelvin}$ and the light intensity variation is less than $1\%$ . Note that the ethanol-water mixing is an exothermic process, which due to the local heating affects the solubility and thus the oversaturation and the light intensity. We took 100 frames to obtain the averaged light intensity of the mixture. The water-adding process lasted until the solution turned opaque enough, after which we obtained the intensity-to-concentration graph, see the black points in Fig. \ref{img:titration}. Based on the data points we performed a $2^{nd}$ order polynomial curve fitting to obtain the titration curve, from which we define the 1$\%$ drop of the light intensity as the point of saturation, as marked by the red star in Fig. \ref{img:titration}. We repeated such titration procedure for 25 different initial compositions of the ethanol-oil mixture, with weight ratios $w_e:w_o$ ranging from $10:1$ to $300:1$, forming the binodal curve as shown in Fig. \ref{img:ternary}. 

Together with the diffusion path, the acquired binodal curve as in Fig. \ref{img:ternary} was then used to quantify the amount of oversaturation, see Fig. \ref{img:binodal}. \citet{Ruschak1972} first formulated the diffusion path theory, which was later applied to the study of nanodroplets formation on surfaces \citep{Lu2017, Li2021}. The diffusion path can be understood as the trajectory of the local fraction of all three phases during the mixing process. As rigorously evaluated in \citet{Li2021}, we assume the diffusion coefficient of water and oil to be the same, meaning that the diffusion path becomes a straight line in the ternary diagram, namely the dashed arrows in Fig. \ref{img:ternary}. For the ease of interpretation, Fig. \ref{img:binodal} can be obtained by coordinate transformation from the water-rich side of Fig. \ref{img:ternary} to a Cartesian coordinate system, mapping the binodal curve and the diffusion path. With the straight-line assumption, the diffusion path depends solely on the initial ethanol-oil ratio. While there is only one binodal curve for all the different compositions, the diffusion path varies with the composition $w_e/w_o$. For a specific water fraction, its saturation concentration is first identified on the binodal curve in Fig. \ref{img:binodal}. Then we connect the saturation point on the binodal curve to the end point of the pure oil phase \citep{Li2021}, at coordinate (0,1), shown as the light green dashed line in Fig. \ref{img:binodal}. The connection between the saturation point and the pure oil phase stems from the consideration that the ethanol-to-water ratio in the mixture remains unchanged in a fluid parcel during oil nucleation \citep{Lu2016}. From the light green dashed line we can extract the length of the segment between the diffusion path and the binodal curve as the oversaturation, displayed as the green solid segment in Fig. \ref{img:binodal}.

Then the oil oversaturation is related to the water concentration as in Fig. \ref{img:oversat}. Fig. \ref{img:oversat} not only displays the significant dependence of the oversaturation on the initial composition, $w_e/w_o$, but also reveals the existence of an upper bound for the oversaturation. The existence of the upper bound might seem counter-intuitive in the first place, as one might expect a monotonic increase with more water introduction. It is, however, not the case in such a ternary liquid system. Further mixing with water beyond the critical fraction makes the nucleated oil re-dissolve into the solution, lowering the oversaturation. While Fig. \ref{img:binodal} already reveals that the segment length seems to decrease at larger water fraction, Fig. \ref{img:oversat} clearly displays this fact, which is also later confirmed by the recorded intensity in the concentration calibration as shown in Fig. \ref{img:calicurve}. The water concentration $C_{\text{water}}$ is the selected value used in the concentration calibration, which will be described below.
\subsection{Light attenuation technique and calibration}

To characterize the concentration of the nucleated oil in the flow, we implemented a light attenuation technique. Unlike the passive scalar studies, the droplet nucleation makes the flow almost opaque, causing the commonly used laser induced fluorescence (LIF) not feasible. We therefore use a light attenuation technique to measure the local concentration. This technique has been widely adapted in turbulent jet and plume studies based on the pioneer work of \citet{CENEDESE1998}, from single phase cases with dye as in \citet{Kikkert2007}, \citet{Allgayer2012}, \citet{Sommeren2012}, and \citet{Mingotti2019b}, to multiphase ones as in \citet{Leppinen2001}, and \citet{Mingotti2015}. The method is based on the Lambert-Beer law, relating the recorded depth-integrated intensity to the concentration field, namely 
\begin{equation}
    \log \left(\frac{I}{I_{\text{ref}}} \right) = -2\gamma Cd,
\label{eq:LA0}
\end{equation}
where $I$ is the recorded intensity of the averaged images, $I_{\text{ref}}$ the reference intensity of the background, $\gamma$ the light absorption property of the dye determined from calibration, $d$ the depth of the substance in the line-of-sight, and $C$ the desired line-of-sight average concentration. The line-of-sight averaged concentration used in the aforementioned papers, however, is not a solution for the nonlinear response we see here. In a pioneering work, \citet{Sutherland2012} analyzed the averaged image based on the axisymmetric discretization, converting the depth-integrated recorded image to the concentration field as a function of radius and height, which serves as a very good strategy for our interests to obtain the oversaturation field as a function of radius and height in the domain.\par

\begin{figure}
\centering
\begin{subfigure}[T]{0.5\textwidth}
\caption{}
\includegraphics[scale=0.5]{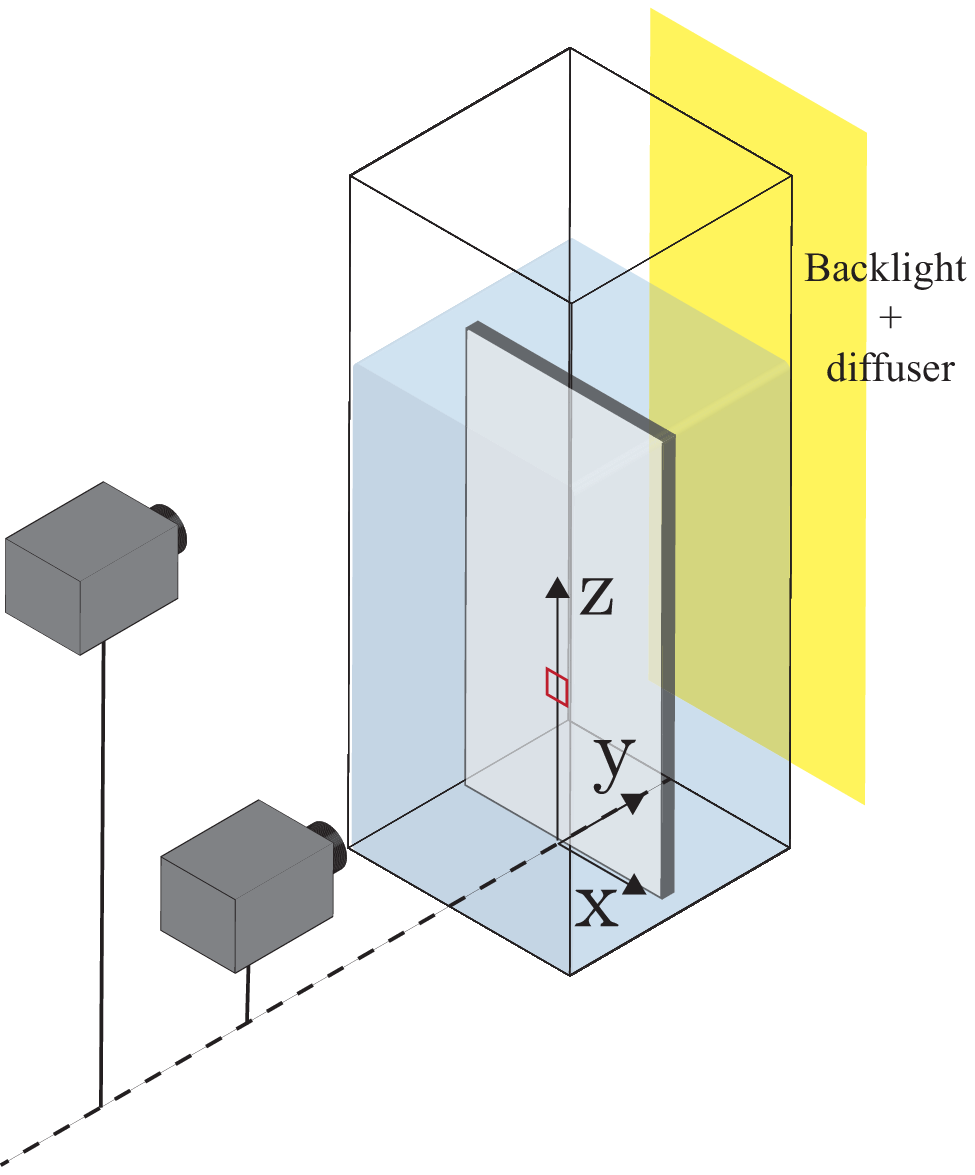}
\label{img:cell}
\end{subfigure}\hspace{-0.1\textwidth}
\begin{subfigure}[T]{0.5\textwidth}
\caption{}
\includegraphics[scale=0.7]{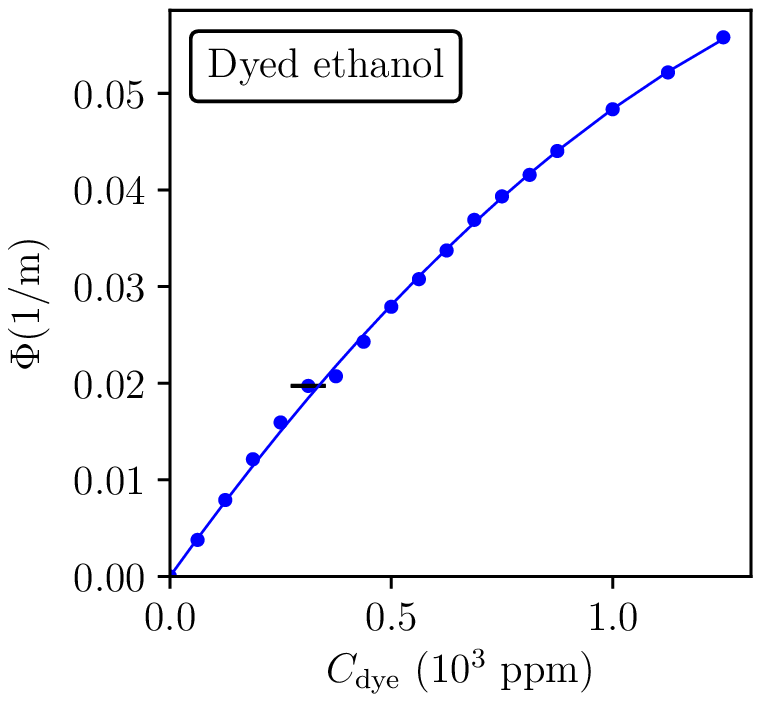}
\label{img:calicurve_dye}
\end{subfigure}%

\begin{subfigure}[T]{0.33\textwidth}
\caption{}
\includegraphics[scale=0.7]{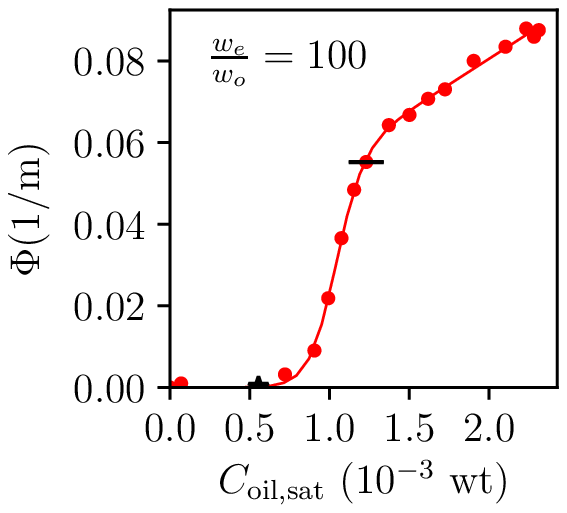}
\label{img:calicurve_100}
\end{subfigure}%
\begin{subfigure}[T]{0.33\textwidth}
\caption{}
\includegraphics[scale=0.7]{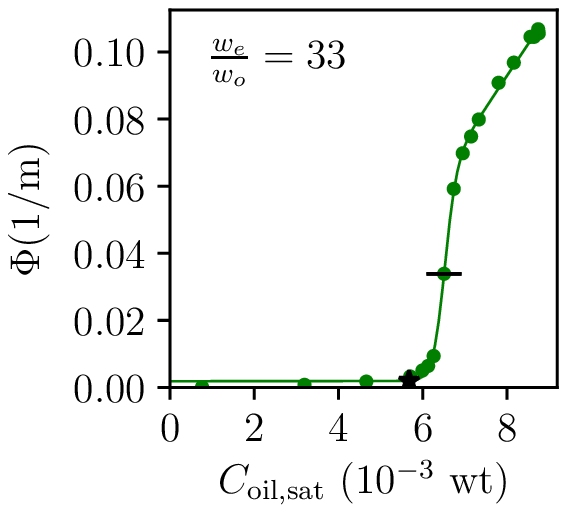}
\label{img:calicurve_33}
\end{subfigure}%
\begin{subfigure}[T]{0.33\textwidth}
\caption{}
\includegraphics[scale=0.7]{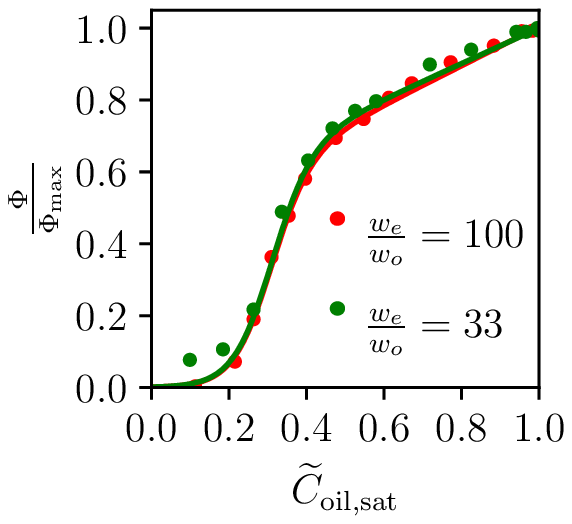}
\label{img:calicurve_all}
\end{subfigure}%

\caption{(a) Calibration cell. The red frame ($x=\ $\SI{0}{\cm}, $z=\ $\SI{10}{\cm}) denotes the selected $5 \times 5$ pixels working unit whose calibration curves are shown in (b)--(e). (b) Calibration curve for the reference dye case. $\Phi$ denotes the degree of light attenuation per unit depth, $\log(\frac{I_{\text{ref}}}{I})/d$, where $d$ is the cell thickness. (c,d) The calibration curve for the two ouzo cases. The abscissa $C_{\text{oil,sat}}$ is the oversaturation of the oil. For (b)--(d) the black error bar represents the standard error of the mean $\Phi$, which is small throughout the calibration. (e) The rescaled calibration curves as in (c), (d).}
\label{img:calicurve}
\end{figure}

We performed our calibration in a \SI{10}{\mm} thick cell as shown in Fig. \ref{img:cell}. The calibration cell is immersed into the tank filled with water, and the illumination and the camera magnification are exactly the same as in the jet experiments, ensuring an in-situ calibration. In spite of the effort to regulate the fluid temperature during the calibration and before the real experiments, the exothermic nature of mixing between water and ethanol in the jet will rise the local temperature. Although the $local$ temperature variation might alter the oversaturation, leading to less accurate calibration, such deviation is simply beyond control or even monitoring during the experiments. With more than \SI{35}{\liter} of water in the ambient and less than \SI{60}{\ml} injected fluid, it is safe to assume that the large volume of water at $\SI{20}{\celsius}\pm \SI{1}{\kelvin}$ is sufficient to mitigate the temperature increase quickly, and wait for future numerical efforts to validate our assumption. As the reference case, the calibration curve for the dyed ethanol is shown in Fig. \ref{img:calicurve_dye}, which was used to calculate the concentration field for the dyed ethanol jet. Red food dye (JO-LA) was added step-by-step followed by taking 200 frames to obtain the averaged light intensity for every $5 \times 5$ pixels working window. Although the linear relation indeed holds within a certain range of dye concentration, we fit a $2^{nd}$ order polynomial to increase its applicability. The nonlinear calibration curve does not appear for the first time. \citet{Sommeren2012} also observed a nonlinear response in their experiment, which they fitted with polynomials, confirming that in-situ calibration also functions well with nonlinear optical response. 

On the other hand, for the ouzo cases, the cell was first filled with a mixture of ethanol and oil at the specific $w_e/w_o$ ratio, and water was then added gradually. Averaging the intensity over 200 frames for every 5 $\times$ 5 pixels working window, the calibration curves as in Fig. \ref{img:calicurve_100} and Fig. \ref{img:calicurve_33} were built for every working window locally in the entire domain, which, however, reveals a completely difference calibration curve from the dye case. The $x$-axis in these two figures, namely the oil oversaturation, were derived from Fig. \ref{img:oversat}. The upper bounds for the ouzo calibration curve in Fig. \ref{img:calicurve_100} and Fig. \ref{img:calicurve_33} are imposed by the theoretical maximum oversaturation shown in Fig. \ref{img:oversat}, namely 0.0023 for $w_e/w_o=100$, and 0.0087 for $w_e/w_o=33$. Such theoretical upper bounds are only valid if the diffusion path does not deviate too much from the straight line assumption as in Fig. \ref{img:binodal}. In the calibration, we did observe that the light attenuation level and the oversaturation decreased as more water was added, see the cluster of data points close to the peak in Figs. \ref{img:calicurve_100} and \ref{img:calicurve_33}. 

\begin{figure}
\centering

\begin{subfigure}[hbt]{0.5\textwidth}
\caption{}
\includegraphics[scale=0.7]{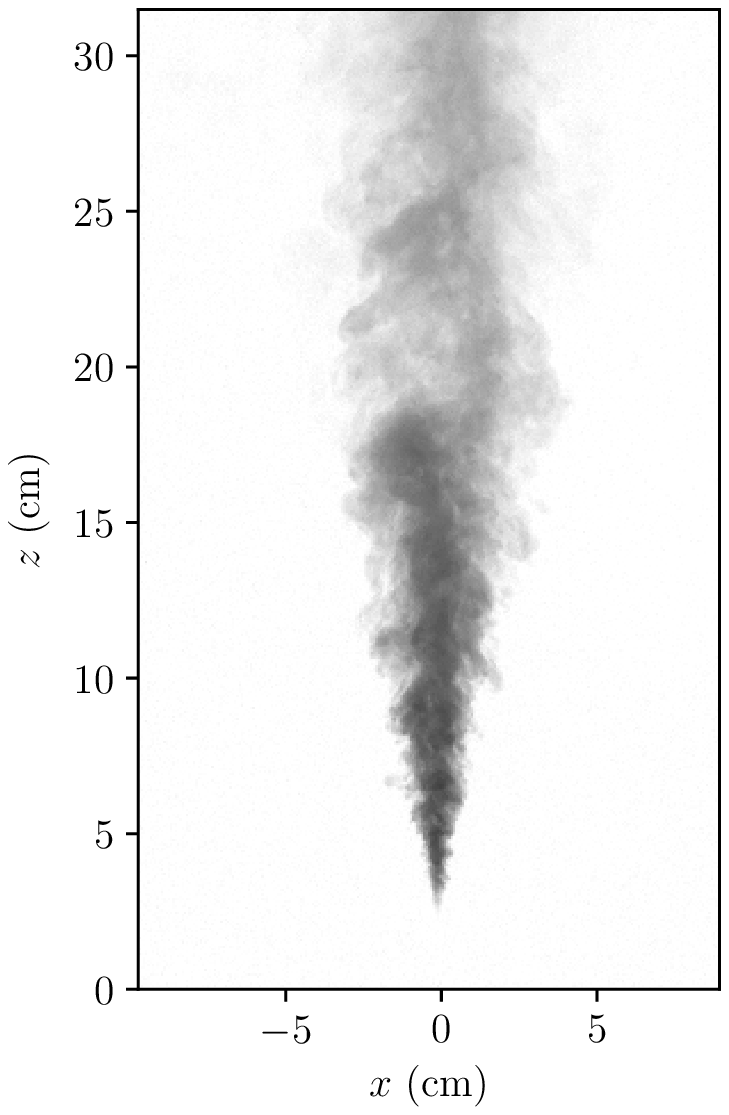}
\label{img:instimg}
\end{subfigure}%
\begin{subfigure}[hbt]{0.5\textwidth}
\caption{}
\includegraphics[scale=0.7]{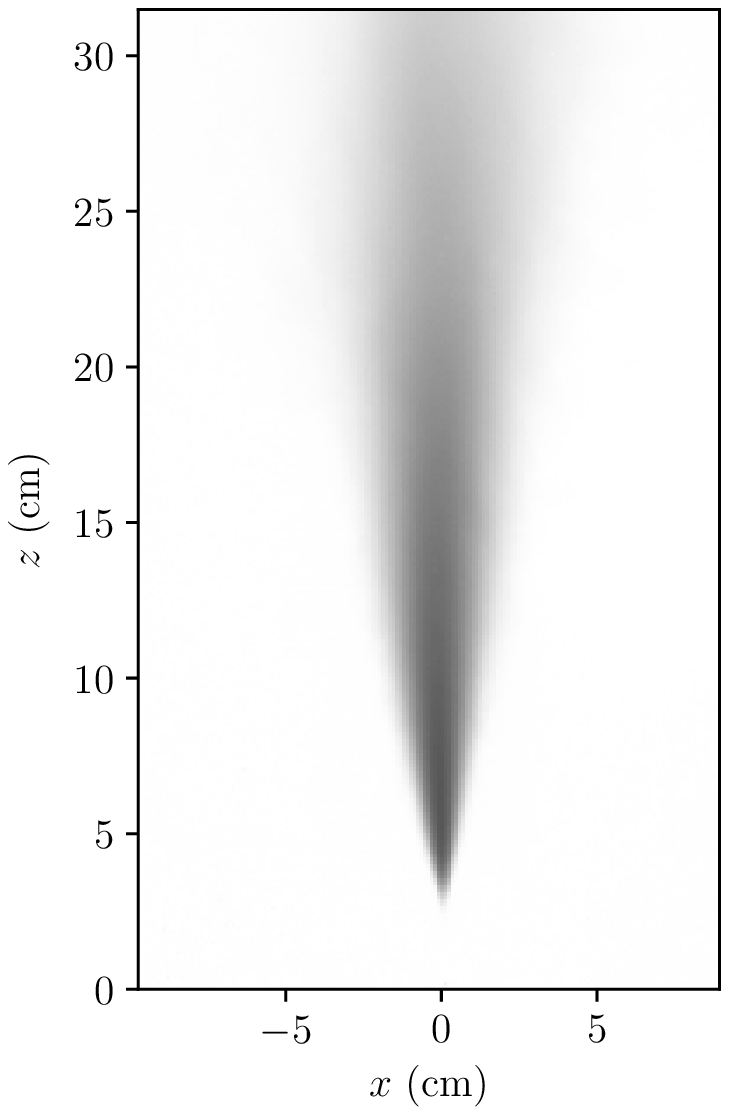}
\label{img:meanimg}
\end{subfigure}\vspace{-2mm} 
\begin{subfigure}[hbt]{0.5\textwidth}
\caption{}
\includegraphics[scale=0.4]{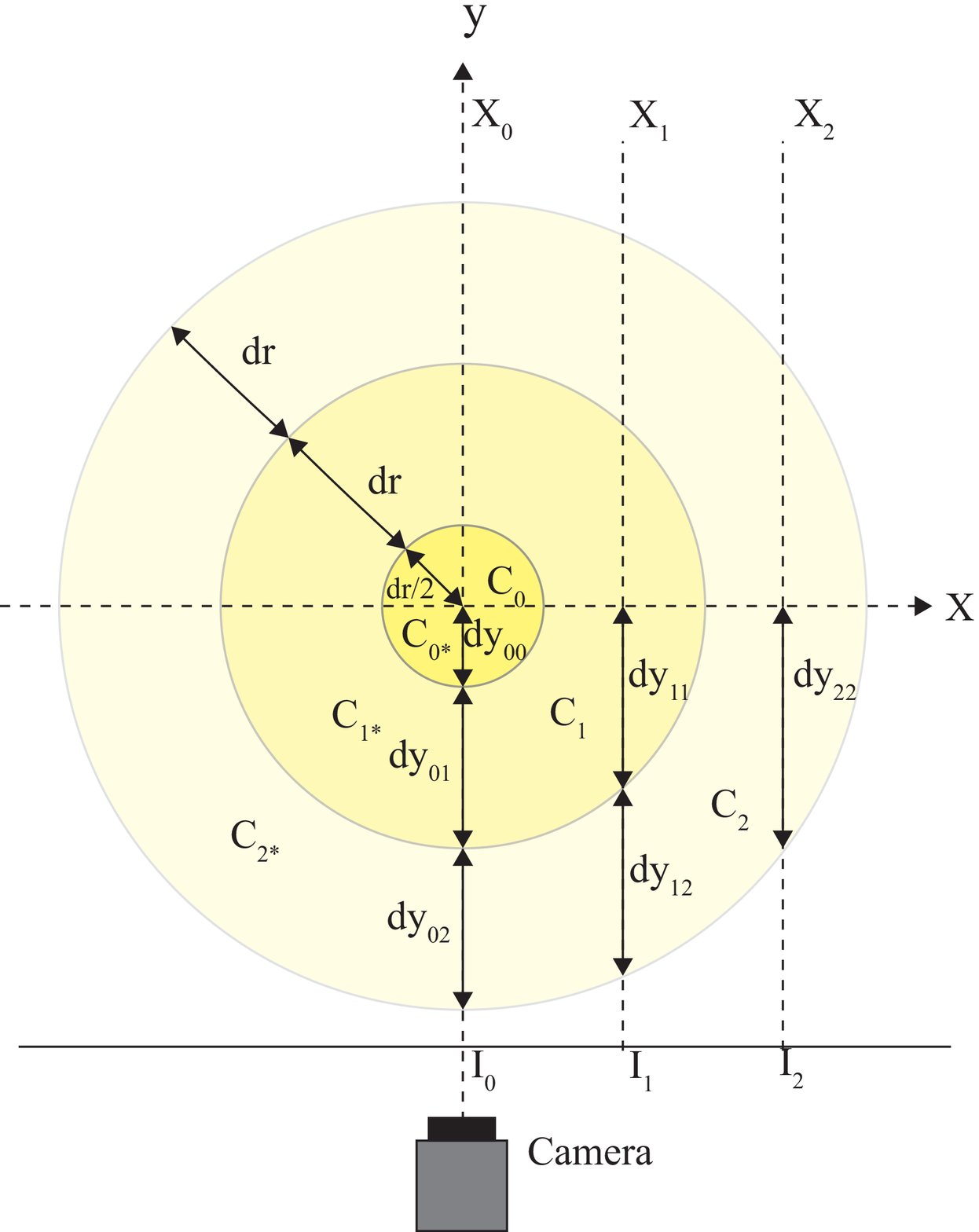}
\label{img:discre}
\end{subfigure}%

\caption{(a) The instantaneous image obtained by dividing the recorded frame by the background frame. (b) The mean image derived by averaging the frames as in (a) over \SI{30}{\second}. $Re_0=1387$ in the image.} (c) An horizontal cross-section of the axisymmetric discretization, similar to that in \citet{Sutherland2012}. Here we do not impose symmetry for the left half ($x<0$) and the right half ($x>0$) of the jet. The averaged image in (b) can be converted to a concentration field by the calibration curve in Fig. \ref{img:calicurve} for every 5 $\times$ 5 pixels, together with the optimization algorithm constructed based on the scheme in (c).
\label{img:imageshow}
\end{figure}

For the case of ouzo, the combination of a logistic function and a linear function fits our calibration data well, written as
\begin{equation}
    \Phi(C) = \frac{\log \left( \frac{I_{\text{ref}}}{I} \right)}{d}= \frac{a_0}{1+a_1e^{a_2(C-a_3)}}+\frac{a_4(C-a_5)}{1+a_1e^{a_2(C-a_3)}},
\label{eq:LA1}
\end{equation}
where $a_0$, $a_1$, $a_2$, $a_3$, $a_4$, and $a_5$ are the fitting parameters, and $C$ is the oil oversaturation, equivalent to the $C_{oil,sat}$ in Fig. \ref{img:calicurve_100}. Note that other fits are possible, but for our purpose we simply need a good approximation of the experimental curve $\Phi(C)$. To compare the calibration curve for the two $w_e/w_o$ ratios, the oversaturation and the attenuation were both rescaled, showing a reasonable collapse, see Fig. \ref{img:calicurve_all}. The abscissa $\widetilde{C}_{\text{oil,sat}}$ in Fig. \ref{img:calicurve_all} is the rescaled oversaturation, which is obtained from  $\widetilde{C}_{\text{oil,sat}} = (C_{\text{oil,sat}}-C_{\text{thresh}})/(C_{\text{max}}-C_{\text{thresh}})$, where $C_{\text{thresh}}$ is the oversaturation where the attenuation reaches 0.1\% of the highest attenuation as indicated by the black star in (c), (d), and $C_{\text{max}}$ is the theoretical upper bound discussed earlier. The rescaled oversaturation also eliminates the flat section of the logistic function for low oversaturation, creating a one-to-one injective map for the calibration. Note that the reason behind such nonlinear optical response is beyond the scope of this research and remains unclear.
Fig. \ref{img:instimg} shows a snapshot of the turbulent jet with oil droplet nucleation, while Fig. \ref{img:meanimg} is the averaged image over 2000--2500 frames as that in Fig. \ref{img:instimg}. The intensity field of a certain height is only included in the averaging operation after the jet has reached that height. The averaged image contains the upstream section only with $z\leq$ \SI{30}{\cm} since the low density dark layer accumulated on top of the tank grew and interfered with the ascending jet, causing the intensity field above $z=$ \SI{30}{\cm} to be averaged relatively poorly (less than \SI{20}{\second}). Fig. \ref{img:discre} shows a horizontal cross-section of the jet, illustrating the discretization scheme similar to \citet{Sutherland2012}, where $dy_{mn}$ can be directly obtained as a function of $dr$ solely based on geometric relations, where $m$ denotes $x$ position on the recorded images, and $n$ the axisymmetric elements in $r$ direction. Note that $m\leq n$ always holds, and we calculate the left half ($x<0$) and right half ($x>0$) of the jet separately, that is $C_n\neq C^{*}_n$. While the concentrations $C_n$ remain as unknowns, the recorded intensity $I_m$ at a specific location from Fig. \ref{img:meanimg} can then be expressed as the integration of the converted intensity calculated from the calibration curve over every discretization element on the light path. 
For the reference dye case with the $2^{nd}$ order polynomial calibration curve, the concentration can be obtained by minimization of $m$ equations with $n$ variables, written as 
\begin{equation}
\log{\left(\frac{I_{ref}}{I_m}\right)} = \sum_{n=0}^{N-1} dy_{mn} \left(a_0  + a_1C_n + a_2C_n^2\right),
\label{eq:LAdye}
\end{equation}

while for the ouzo cases with calibration curve as Eq. (\ref{eq:LA1}), the equations for minimization are written as
\begin{equation}
\log{\left(\frac{I_{ref}}{I_m}\right)} = \sum_{n=0}^{N-1} dy_{mn} \left( \frac{a_0}{1+a_1e^{a_2(C_n-a_3)}}+\frac{a_4(C_n-a_5)}{1+a_1e^{a_2(C_n-a_3)}}\right)
\label{eq:LAouzo}
\end{equation}
The calculation procedure is repeated for every height, leading to the oversaturation fields as in Fig. \ref{img:cmap}. We detail the procedures of the optimization process and the evaluation of the results in Appendix A.

We add a note of care: The superposition of attenuation level across the layers of axisymmetric unit illustrated in Fig. \ref{img:discre} might be questionable for a scattering medium. The summation in eq. (\ref{eq:LAouzo}) is fully applicable if the level of attenuation is proportional to the depth of the medium; however, such algorithm does not hold when the light propagates through a scattering medium. Despite that our scattering is relatively low, the proposed method and the obtained results in this paper should be further evaluated in the future. We will discuss this issue in more detail in Appendix C.

\subsection{Droplet size detection}
\begin{figure}
\centering
\begin{subfigure}[t]{0.33\textwidth}
\caption{}
\includegraphics[scale=0.8]{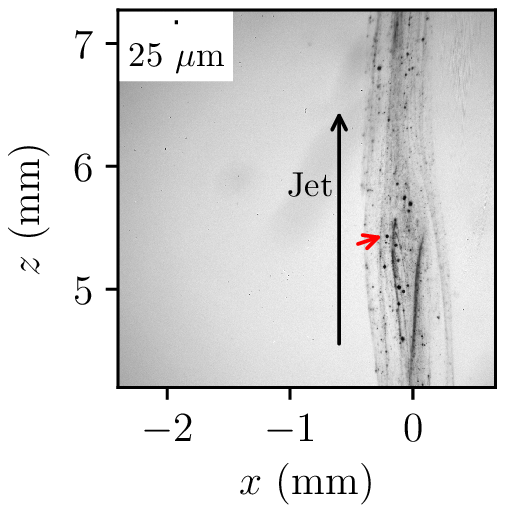}
\label{img:nucslow}
\end{subfigure}%
\begin{subfigure}[t]{0.33\textwidth}
\caption{}
\includegraphics[scale=0.8]{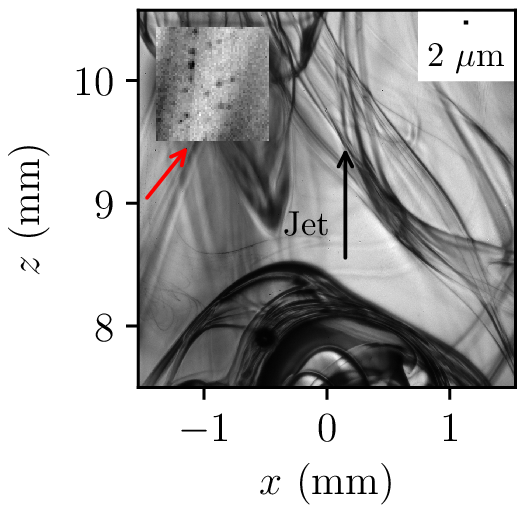}
\label{img:nuc100}
\end{subfigure}%
\begin{subfigure}[t]{0.33\textwidth}
\caption{}
\includegraphics[scale=0.8]{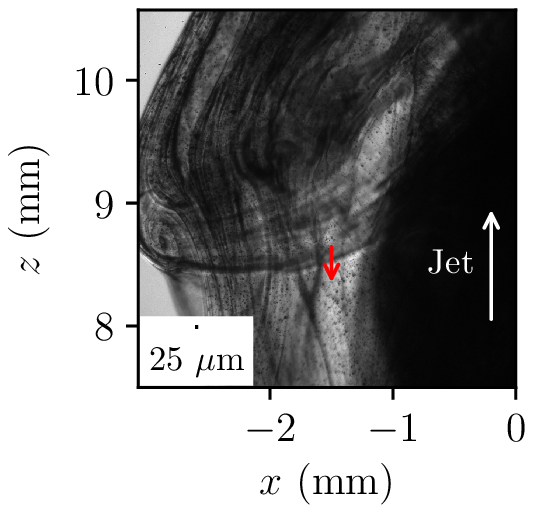}
\label{img:nuc33}
\end{subfigure}%

\caption{Identification of nucleated droplets. For comparison, we show a droplet of 25 $\mu$m diameter in all three cases. (a) The nucleated droplets for $w_e/w_o=100$ when the injection just stopped (i.e.,$u_z$ approached $0$). (a) serves as a reference laminar case where the droplet nucleation can be easily identified. (b) The nucleated drops of a jet with $w_e/w_o=100$ in the turbulent regime for $Re_0 = 555$. The black thin sheets are the fluid interfaces due to refraction, while the black dots in the inset are the nucleated droplets with 17 $\times$ magnification, which are obviously smaller than $25$ $\mu$m. (c) The nucleated droplets for $Re_0 = 555$ and $w_e/w_o=33$, which are larger and easier to identify}
\label{img:nucleation}
\end{figure}

The nucleated droplets through solvent exchange are generally referred to as micro- or even nano-sized droplets. However, experimentally measured sizes are based on cases with no flow (\citet{Tan2019}) or laminar flow \citep{Hajian2015, Lu2015}. To measure the sizes of the nucleated droplets in the case of a turbulent jet, we utilize a Navitar 12 $\times$ objective on a Photron FASTCAM Mini AX200 camera, achieving a resolution of $3$ $\mu$m/pixel, see Fig. \ref{img:nucleation}. Fig. \ref{img:nucslow} clearly shows the nucleated droplets in the plume after the injection stopped, which is similar to the laminar flow scenario in the aforementioned literature.  Comparing Figs. \ref{img:nuc100} and \ref{img:nuc33}, we see that the droplets were indeed micro-sized ($\mathcal{O}(10$ $\mu$m)) for both turbulent cases. The droplets for $w_e/w_o=33$ are larger and thus easier to identify, which is in line with the findings of \citet{Vitale2003}.

\subsection{PIV}
While the light attenuation technique resolved the mean oversaturation field, we relied on particle image velocimetry (PIV) to measure the velocity field. Ethanol seeded with 20--\SI{50}{\micro\metre} fluorescent particles was injected into the water-filled tank, which was also seeded with the same particles in order to prevent measuring a velocity that is solely based on entrained or injected liquid. Note that we used the obtained velocity field of the ethanol jet to represent that of the ouzo jet, assuming that the micro-sized nucleated oil droplets can be considered as tracer particles which barely affect the velocity field. The Photron cameras were replaced with LaVision Imager sCMOS cameras with resolution 2560 $\times$ 2160 pixels, capturing the image pairs on the \SI{1}{\mm} thick laser sheet created by a dual cavity laser (Quantel Evergreen 145 laser, \SI{532}{\nm}). The velocity fields are calculated using the multi-pass method with a starting window size 64 $\times$ 64 pixels to a final size 16 $\times$ 16 pixels with 25$\%$ overlap.

Considering that the velocity decreases drastically as the jet travels downstream, the experiments were performed with two different time intervals $dt$ between image-pairs to resolve both the far field and the near field, followed by averaging over 1000 frames to obtain the mean velocity fields. Note that the region below $z=$ \SI{3}{\cm} cannot be resolved because the jet is thinner than the \SI{1}{\mm} laser thickness. 

\begin{figure}
\centering
\begin{subfigure}[t]{0.5\textwidth}
\caption{}
\includegraphics[scale=0.7]{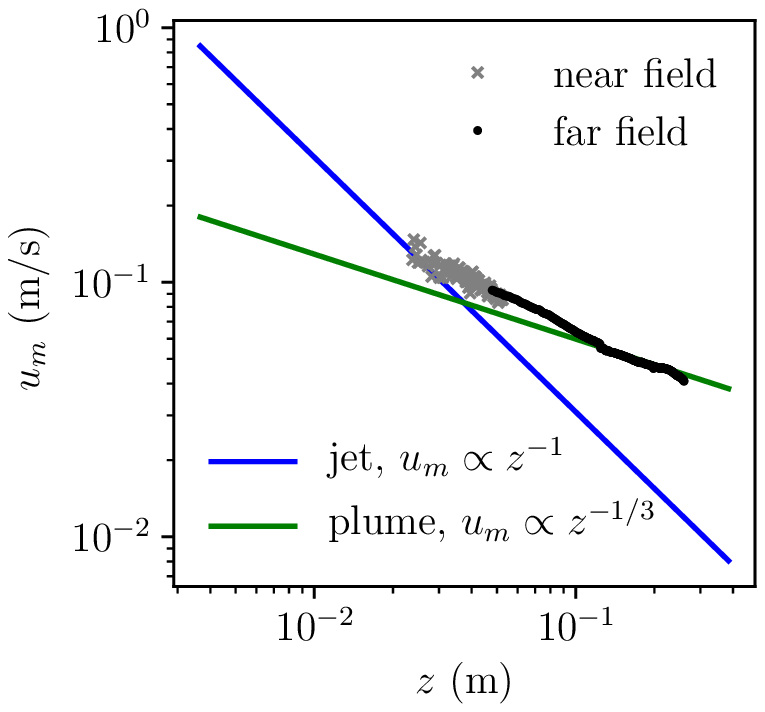}
\label{img:um_q20}
\end{subfigure}%
\begin{subfigure}[t]{0.5\textwidth}
\caption{}\hspace{-2mm}
\includegraphics[scale=0.7]{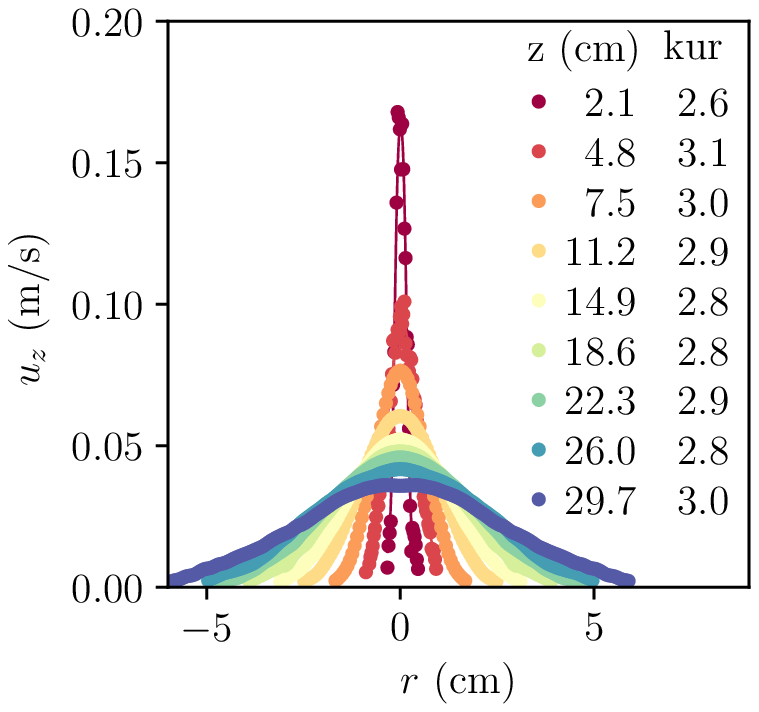}
\label{img:piv_sig20}
\end{subfigure}%

\begin{subfigure}[t]{0.5\textwidth}
\caption{}
\includegraphics[scale=0.7]{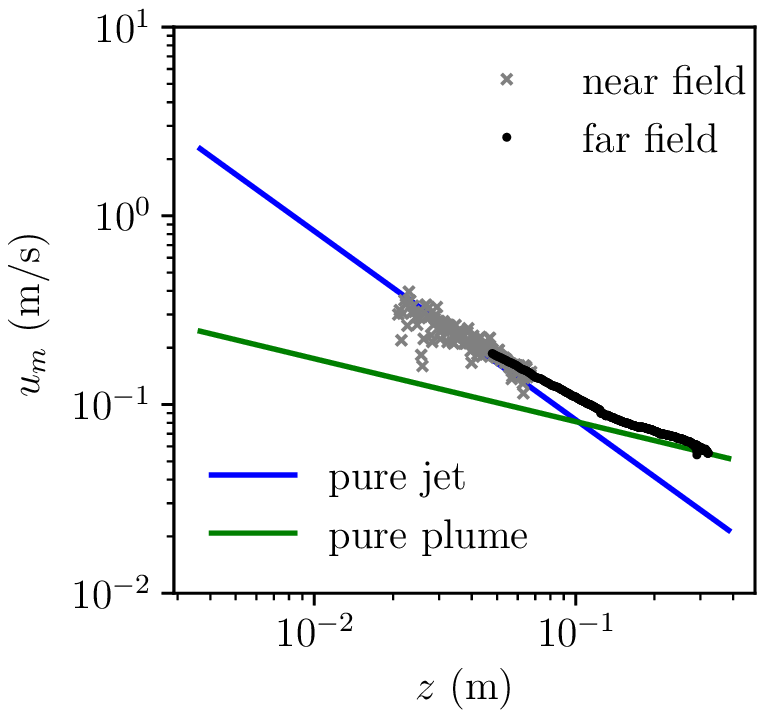}
\label{img:um_q50}
\end{subfigure}%
\begin{subfigure}[t]{0.5\textwidth}
\caption{}
\includegraphics[scale=0.7]{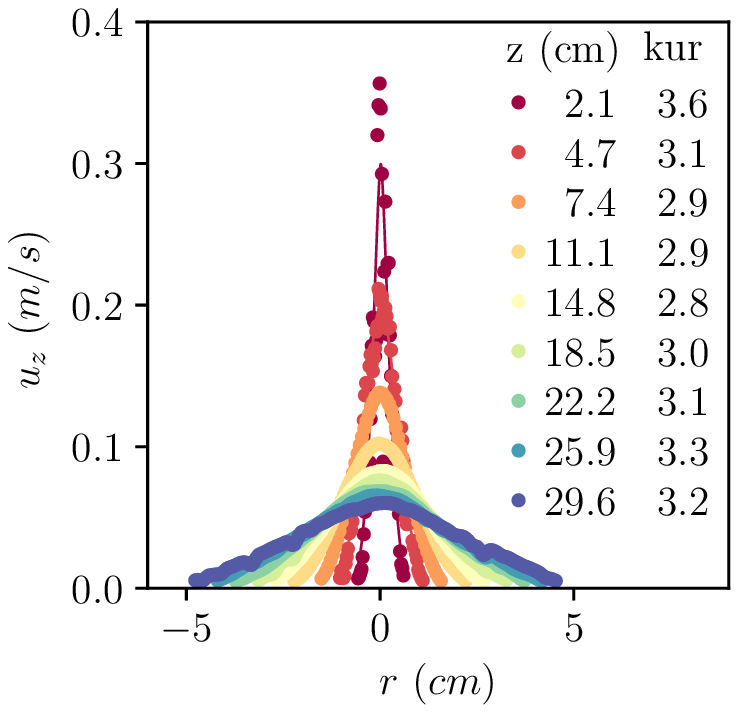}
\label{img:piv_sig50}
\end{subfigure}%
\caption{The mean velocity field obtained by PIV measurements at (a,b) $Re_0=555$, and (c,d) $Re_0=1387$ . (a,c) The centerline velocity in $z$-direction, $u_m$, from the theoretical model and the valid measurements of the both fields. (b,d) The measured and fitted radial profile of the streamwise velocity $u_z$. Kur in the legend is abbreviation for kurtosis.}
\label{img:piv}
\end{figure}

Obtained from the velocity field, Figs. \ref{img:um_q20}, \ref{img:um_q50} show the mean centerline evolution of velocity in $z$-direction, $u_m$. Theoretical models prediction for a pure jet and a pure plume are also presented:
\begin{align}
& u_{\text{m,jet}}= 4.2M^{1/2}z^{-1}, \nonumber\\
& u_{\text{m,plume}}= 3.2B^{1/3}z^{-1/3},
\label{eq:jp}
\end{align}
where $M=Q^2/(\frac{1}{4}\pi d^2)$ is the initial momentum flux, and $B=Qg(\rho_{\text{jet}}-\rho_{\text{amb}})/\rho_{\text{amb}}$ is the initial buoyancy flux, see the caption in Table \ref{tbl:condition}.

The coefficients obtained from our measurements are lower than the values in the fully turbulent jet and plumes, which are 5--7.5 for jet and 3.4--3.9 for plume \citep{Fischer1979,List1982}. We attribute such difference to the relatively low Reynolds number in this study, where the jet regime lies right after the laminar-to-turbulent transition.

Such comparison with the models reveals that the initial ethanol jet quickly transitioned toward a plume under strong buoyancy effect. Figs. \ref{img:piv_sig20}, \ref{img:piv_sig50} show the radial flow profile of the streamwise velocity $u_z$, which are later used to calculate the oversaturation flow rate as in Fig. \ref{img:flux}.

\section{Oversaturation field}
\begin{figure}
\centering
\begin{subfigure}[t]{1\textwidth}
\includegraphics[scale=0.7]{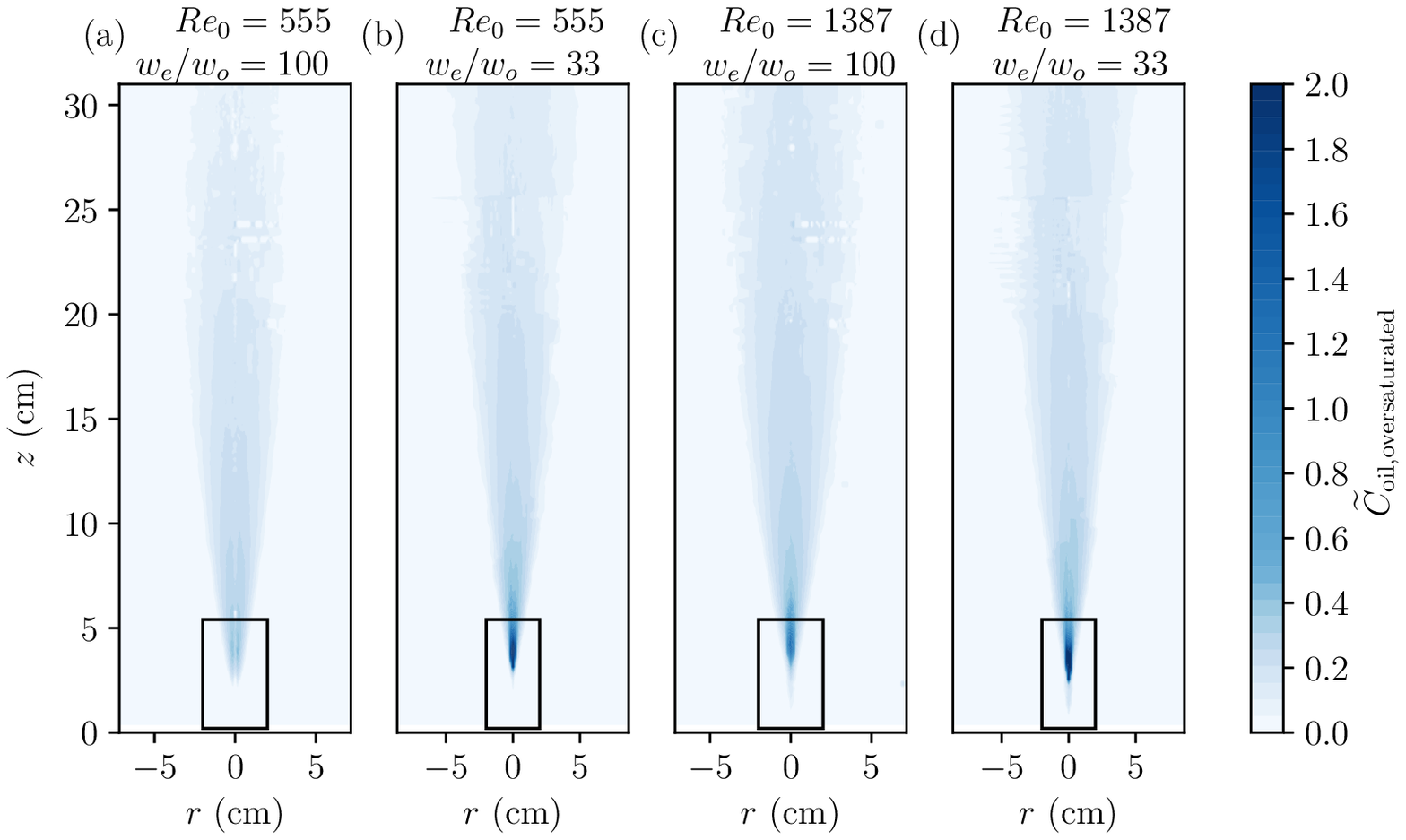}
\label{img:cmapfar}
\end{subfigure}

\begin{subfigure}[t]{1\textwidth}
\centering \hspace{-4mm}
\includegraphics[scale=0.7]{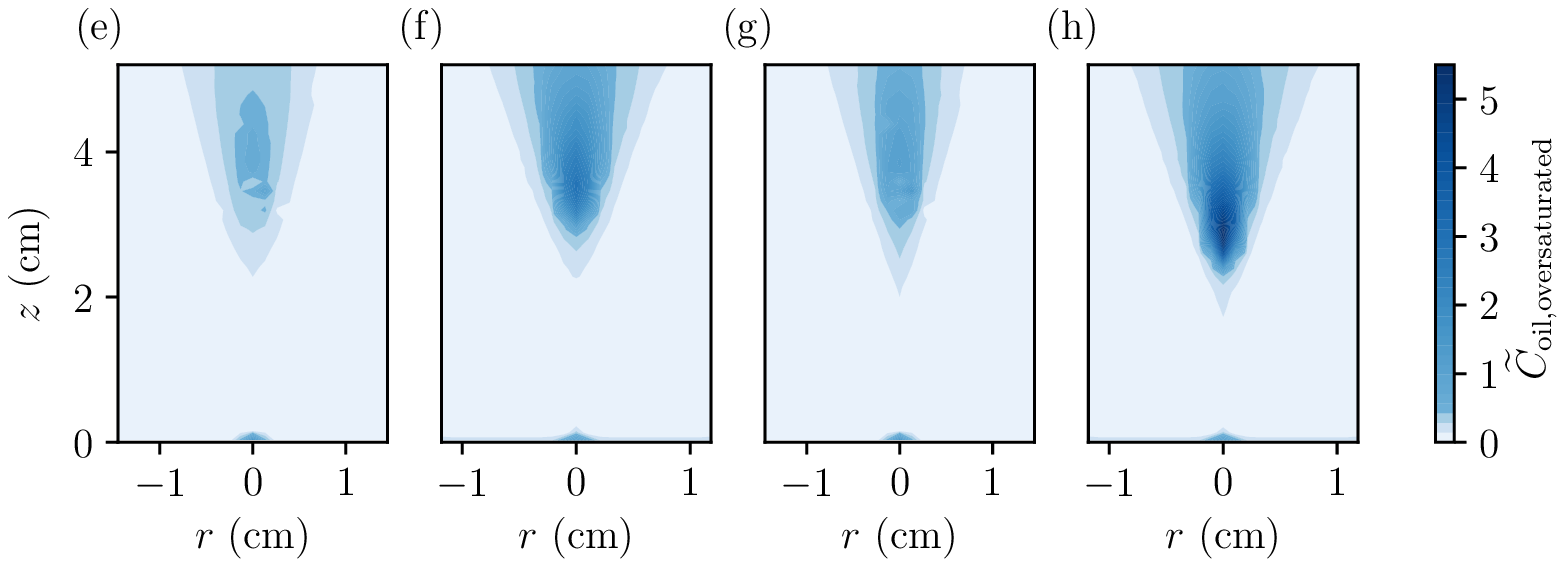}
\label{img:cmapnear}
\end{subfigure}
\caption{Calculated rescaled oversaturation fields for different $Re_0$ and $w_e/w_o$. Panels (a--d) show the results derived from the far view camera, while (e--h) exhibit the results obtained from the zoomed-in camera, which correspond to the black square in (a--d), respectively, with the same color scheme below $\widetilde{C}_{\text{oil,oversat}}$=1. The white spots in the contour map are caused by the local minima found during optimization, which poses little effect on the analysis in \S4 as we exclude those defects for curve-fitting. Note that the abscissa is converted from $x$ to $r$ using the axisymmetric transformation in Fig. \ref{img:discre}}.
\label{img:cmap}
\end{figure}

Following the procedure detailed in \S2.3, the optimization problem was solved to construct the oversaturation field, Fig. \ref{img:cmap}. As shown in Figs. 3a--3d, the calculated $\widetilde{C}_{\text{oil,oversat}}$ in most of the domain fall below 1. Such results suggest that the aforementioned procedures in \S2 work decently, including titration, straight diffusion path assumption, calibration, and the optimization. As for the region very close to the needle in the near field, the local oversaturation exceeds the theoretical upper bound, leading to $\widetilde{C}_{\text{oil,oversat}}>1$. While the actual cause of such results remains uncertain, we believe that the most probable cause is the violation of the straight diffusion path assumption in that region. Such violation might have resulted from the intense mixing and cross-component diffusion upon the onset of turbulence in the initial shear layer.

As the far view results at the initial shear layer might be biased by the resolution constraint, the results obtained from zoomed-in recordings are presented in Figs. 3e--3h, corresponding to the regions marked by the black frame in Figs. 3a--3d. Although the oversaturation might not be quantitatively correct in this region, the zoomed-in view results further confirm the intense nucleation in the initial shear layer. In the study of aerosol formation in turbulent gas jets, \citet{Lesniewski1998} also pointed out that the nucleation of the target substance is largely confined to the initial shear layer, especially when the vapor mole fraction of the target substance is low. While the system of the current work is different from that of the aerosol study, the two processes both couple the variation of thermodynamic states and ambient turbulent shear flow, leading to the nucleation of droplets or particles.

Considering the opaque nature of the ouzo jet flow, the calculated results cannot be verified by other concentration detection methods such as LIF. Bearing the limitation in mind, what we can do is to check the consistency of the results by repeatability tests and comparing between the far view and the  zoomed-in view results obtained from two different cameras. While the repeatability will be demonstrated in \S4, Fig. \ref{img:fn} in Appendix B shows a quantitative comparison between the two recordings using the centerline evolution and the radial profile. As detailed in Appendix B, for region with $\widetilde{C}_{\text{oil,oversat}}\leq 1$, the proposed method delivers consistent results between the far view and zoomed-in view recordings.

Note that the turbulence started to develop after a certain laminar length (\citet{Hassanzadeh2021}), which is subtracted to obtain the virtual origin in $z$-direction for the analysis in \S4. The concentration fields for the dyed-ethanol injection were also obtained by the same method, which are not shown here but will be exploited as reference cases in the analysis in \S4.

\section{Results}
\begin{figure}
\hspace{-2mm}
\centering
\includegraphics[scale=0.7]{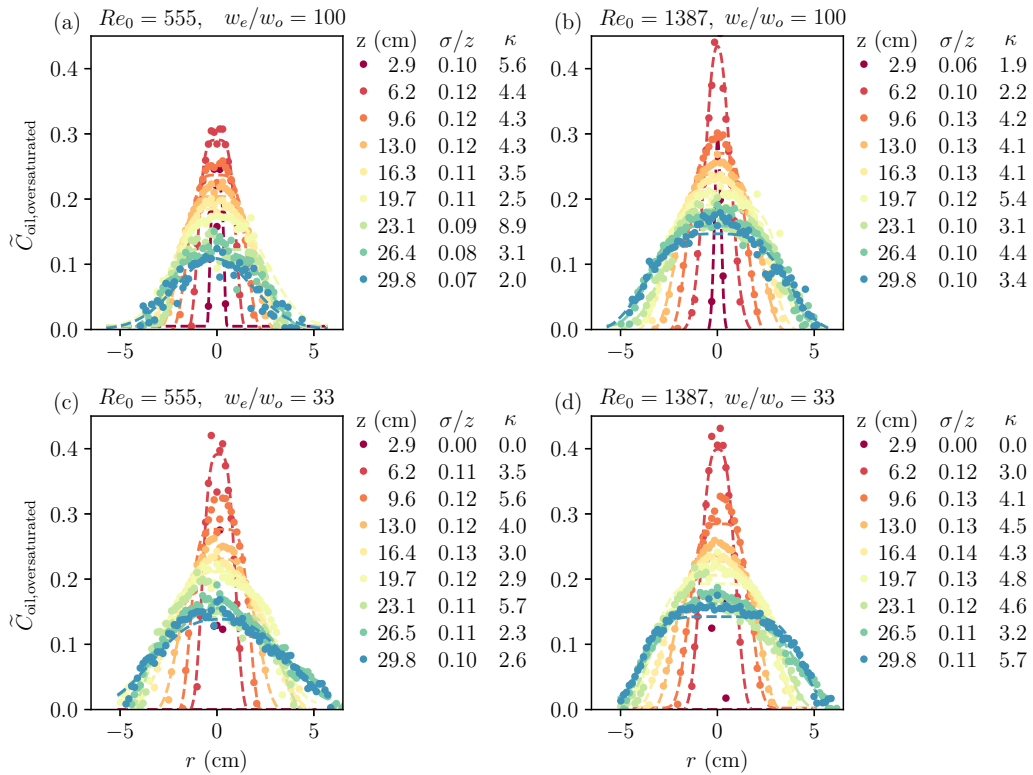}
\caption{The radial oversaturation profiles for different $Re_0$ and $w_e/w_o$. The dots are the data points and the dashed lines are curves fitted with Eq. (\ref{eq:fit}). The table next to the figure gives the obtained fitting parameters $\sigma/z$ \& $\kappa$ for a variety of heights.}
\label{img:radial}
\end{figure}

\subsection{Generalized normal distribution}

To better capture the radial profile of oversaturation field in Fig. \ref{img:cmap}, we fitted the results by a generalization of a normal distribution, 
\begin{equation}
C(r) = C_m e^{-\frac{|r|^{\kappa}}{\kappa \sigma^{\kappa}}} + C_0,
\label{eq:fit}
\end{equation}
where $C_m$ is the centerline oversaturation, where $\sigma$ is the width of the jet, $\kappa$ the parameter to characterize the potential non-Gaussian profile, and $C_0$ the background offset of the oversaturation, which is also very close to zero. When $\kappa = 2$ the profile becomes the standard Gaussian case. Fig. \ref{img:radial} showcases the quality of the fitting, where the center received less weight than the outer region based on the amount of information contained, that is, the area of the corresponding discretized annulus region in Fig. \ref{img:discre}. Also, we exclude the data at any specific height with less than 5 data points to optimize the quality of the fitting. The fitted variables and the fitted profile are then used in the following sections to analyze the evolution of the centerline oversaturation and the jet profile. 

As discussed, $\widetilde{C}_{\text{oil,oversat}}>1$ in Fig. \ref{img:cmap} suggests that the data there is less reliable. We hope to obtain the profile and centerline value only based on the reliable data points, Therefore, data points with $\widetilde{C}_{\text{oil,oversat}}>1$ are also excluded from curve-fitting, but larger fitted $\widetilde{C}_{\text{oil,oversat}}$ is allowed. Such fitting strategy somehow affects the centerline value in the near field, where $\widetilde{C}_{\text{oil,oversat}}>1$. We present and discuss the difference in centerline value between the minimization and the following curve-fitting in Appendix B. The current method cannot work properly in the initial shear layer; we would like to remind the readers that the results in that region cannot be interpreted by our curve-fitting procedures nor our minimization procedure, we therefore should keep this in mind when interpreting the analysis in \S4. 
\subsection{Radial profile}

\begin{figure}
\centering
\hspace{-0.03\textwidth}
\begin{subfigure}[t]{0.42\textwidth}
\caption{}
\includegraphics[scale=0.8]{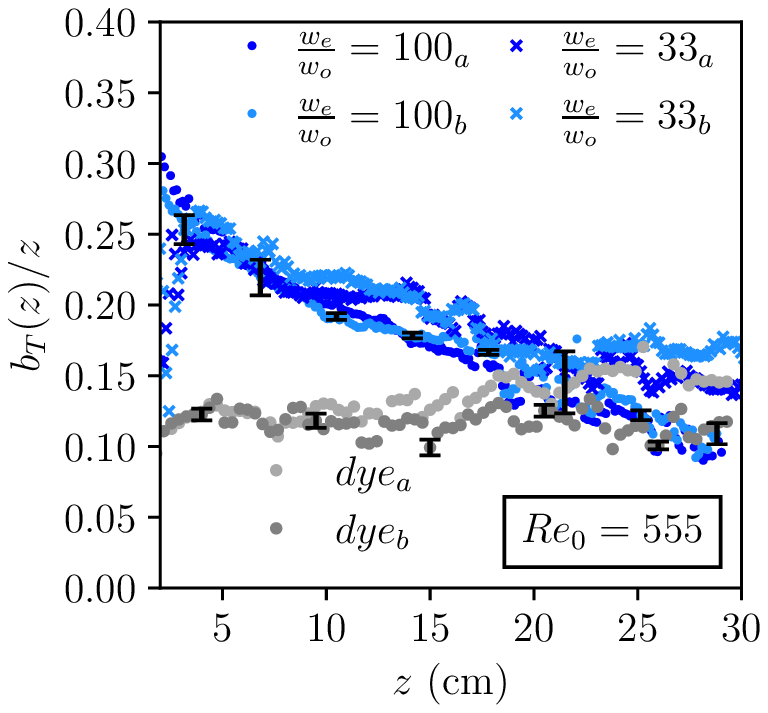}
\label{img:spreadq20}
\end{subfigure}
\hspace{0.08\textwidth}
\begin{subfigure}[t]{0.45\textwidth}
\caption{}
\includegraphics[scale=0.8]{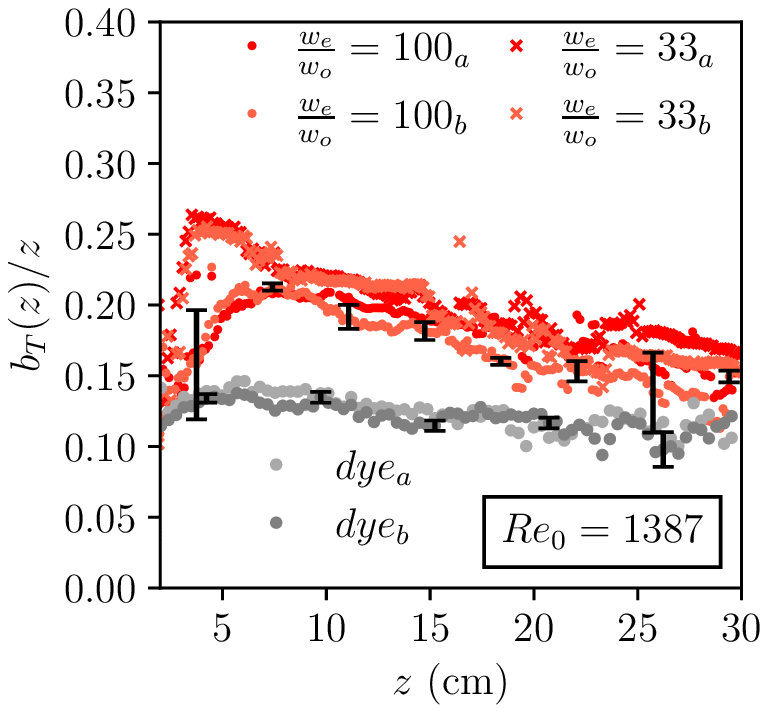}
\label{img:spreadq50}
\end{subfigure}\vspace{-5mm}\hspace{-5mm}
\begin{subfigure}[t]{0.42\textwidth}
\caption{}
\includegraphics[scale=0.8]{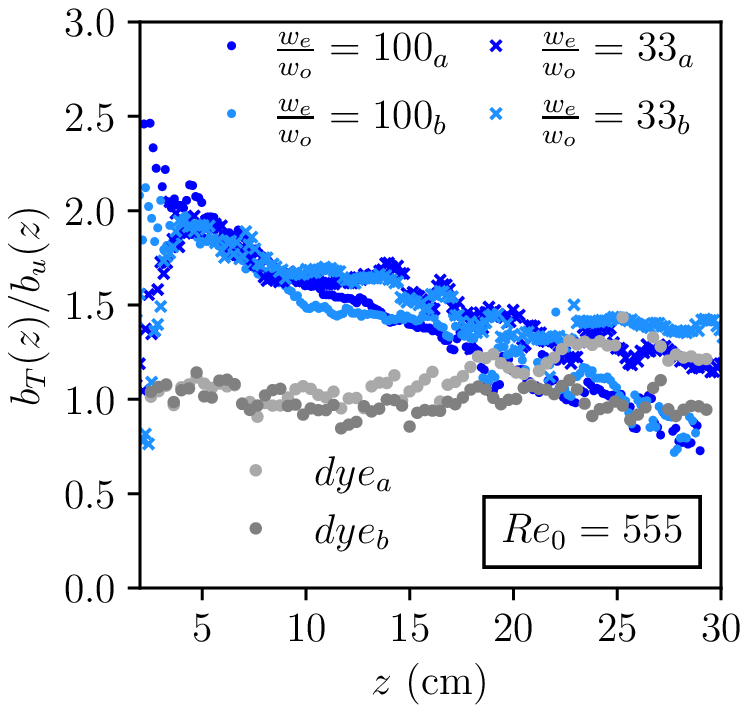}
\label{img:btbuq20}
\end{subfigure}
\hspace{0.08\textwidth}
\begin{subfigure}[t]{0.45\textwidth}
\caption{}
\includegraphics[scale=0.8]{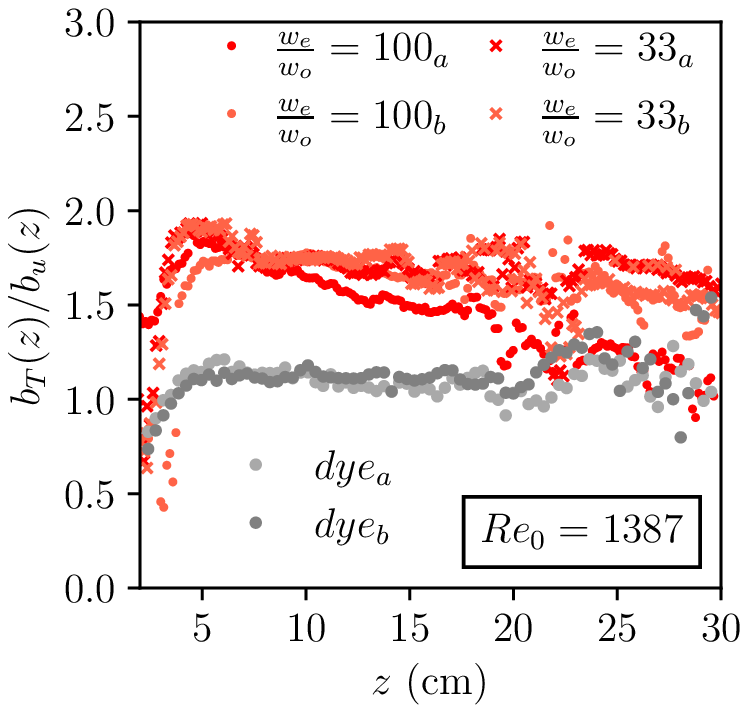}
\label{img:btbuq50}
\end{subfigure}%
\caption{The radial oversaturation profile of the buoyant jets. The errorbars represent the uncertainty of the curve fitting discussed in \S4.1. Panels (a) and (b) show the spatial evolution of the jet width ($b_T(z)$) divided by the vertical distance $z$ by probing the radial position where the concentration reach 1/e of the fitted peak. For each parameter set we show the results of two repeated experiments, denoted by subscripts a, and b. (c,d) display the jet width ratio between the oversaturation (concentration) profile $b_T$ and the velocity profile $b_u$.}
\label{img:radial}
\end{figure}

In the classic jet and plume studies, the jet width of the concentration field $b_T(z)$ is usually determined by the position where the value drops to $1/e$ of the centerline value. For a self-similar flow, $b_T(z)/z$ is a constant at around $0.127$ for pure jet, and $0.12$ for pure plume \citep{Fischer1979}. When the radial flow profile is Gaussian, the jet width determined from the definition above, $b_T(z)$, and $\sigma$ obtained from Eq. (\ref{eq:fit}) are the same. However, from Fig.\ref{img:radial} we can tell the radial profile is non-Gaussian for the ouzo cases, therefore we implemented the traditional definition of $b_T(z)$ to characterize the jet width and plot the evolution of $b_T(z)/z$ in Figs. \ref{img:spreadq20} and \ref{img:spreadq50}. While the reference dye cases match well with the constant value reported in the literature \citep{Fischer1979}, the ouzo cases exhibit a continuously decreasing trend after the initial shear layer, starting with twice the value of the reference case and approaching the reference value as the jet evolved downstream. The larger jet width suggests a wider spread of the nucleation front, following the turbulent entrainment and mixing at the meandering turbulent/non-turbulent interface (TNTI) \citep{Westerweel2009,Watanabe2015a}. Note that we study the averaged density profile, nucleation visible in instantaneous images is very likely to not be in the center and continuously moving around. \citet{Neuber2017} numerically investigated the nucleation and condensation of aerosol in gas turbulent jets, showing that the largest supersaturation level and nucleation rate are both not located in the centerline. While we do believe that the nucleation is most intense at the TNTI, the radial dispersion of the droplets nucleated there and the streamwise dispersion of the dense droplets nucleated in the initial shear layer contribute to the relatively higher centerline value in the mean profile. As the oil, originally dissolved in the injected fluid, nucleates and gradually gets consumed with increasing $z$ the ratio $b_T(z)/z$ also approaches the reference dye case. Note that the spatial evolution of $b_T(z)/z$ is nearly independent of $Re_0$ and $w_e/w_o$ upstream, while the dependence appears downstream. Higher initial oil fraction and higher $Re_0$ can both extend the wider profile further downstream.

Figs. \ref{img:btbuq20} and \ref{img:btbuq50} show the jet width ratio between the oversaturation profile and the velocity profile. The ratio for the reference cases reveals that the concentration spreads slightly wider than the momentum, namely $b_T/b_u$ slightly larger than 1, which is consistent with the literature value. For the ouzo jets, the jet width ratio is roughly two times larger upstream, and gradually decreases as the flow evolves downstream.

\begin{figure}
\vspace{0.03\textwidth}
\centering
\hspace{-0.05\textwidth}
\begin{subfigure}[t]{0.4\textwidth}
\caption{}
\hspace{0.06\textwidth}
\includegraphics[scale=0.8]{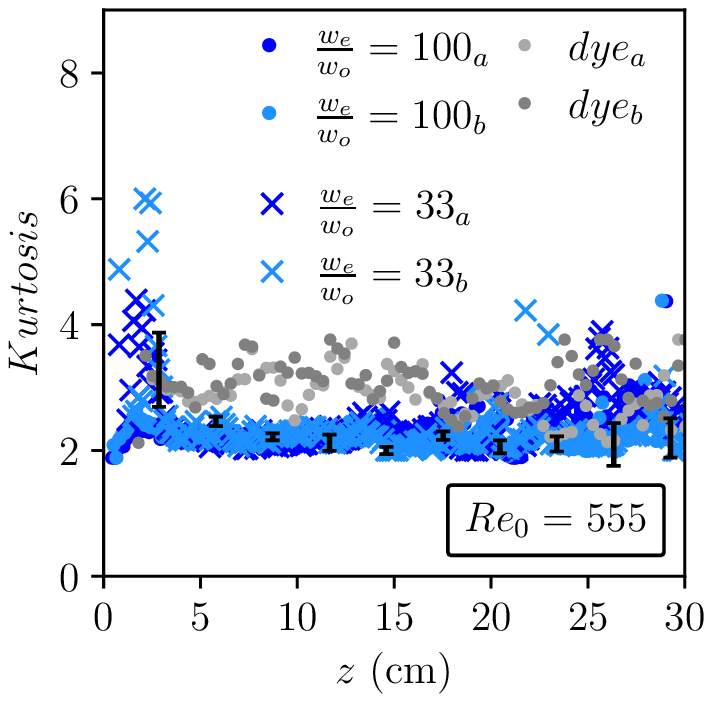}
\label{img:kurq20}
\end{subfigure}
\hspace{0.094\textwidth}
\begin{subfigure}[t]{0.485\textwidth}
\caption{}
\hspace{0.06\textwidth}
\includegraphics[scale=0.8]{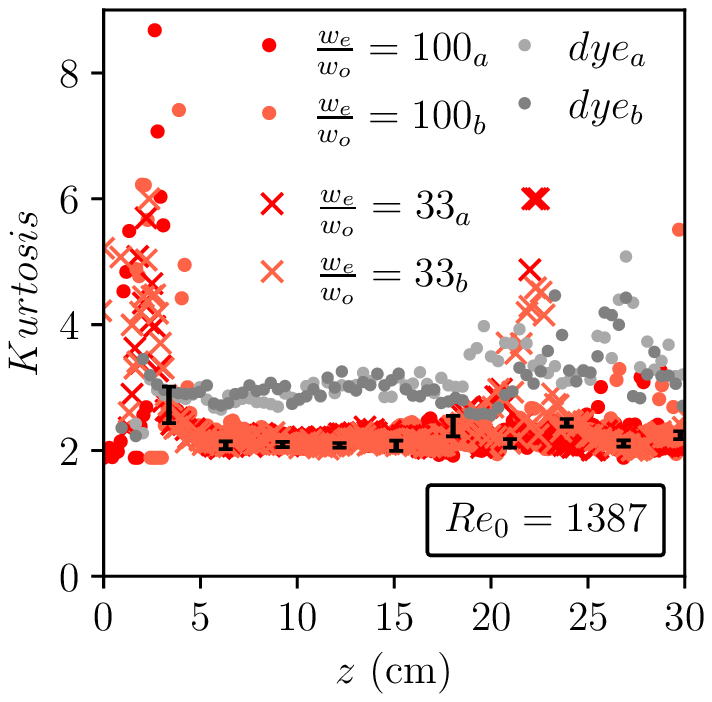}
\label{img:kurq50}
\end{subfigure}%
\caption{Kurtosis calculated from the fitted parameter $\kappa$ for (a) $Re_0=555$ and (b) $Re_0=1387$}
\label{img:kur}
\end{figure}
As the ratio $b_T(z)/z$ for the ouzo jet is larger than the dye case, we wonder how the nucleation changes the distribution of the radial profile. To characterize the radial profile, we calculate the spatial evolution of the kurtosis from the fitted parameter $\kappa$ of Eq. (\ref{eq:fit}), see Figs. \ref{img:kurq20} and \ref{img:kurq50} for the results. We can interpret the concentration profile in Eq. (\ref{eq:fit}) as a probability density function of the concentration $\text{PDF}(C)$ for which we find that $C_m$, $\sigma$, and $\kappa$ are constraint in such a way that  $\int \nolimits_{-\infty}^\infty \text{PDF}(C)dr=1$, as per the definition of the PDF. From that we can express $C_m$ in terms of $\kappa$ and $\sigma$. The kurtosis can then be calculated from its definition in terms of the 2${}^{\text{nd}}$ and 4${}^{\text{th}}$ central moments:

\begin{align}
\text{PDF}(C) &= \frac{\kappa ^{-1/\kappa }}{2 \sigma  \Gamma \left(1+\frac{1}{\kappa }\right)} e^{-\frac{| r| ^{\kappa }}{\kappa \sigma^\kappa}}\\
\text{kurtosis} &= \frac{\int \limits_{-\infty}^\infty r^4 \text{PDF}(C) dr}{\left( \int \limits_{-\infty}^\infty r^2 \text{PDF}(C) dr \right)^2} = \frac{\Gamma \left(\frac{1}{\kappa }\right) \Gamma \left(\frac{5}{\kappa }\right)}{\Gamma \left(\frac{3}{\kappa }\right)^2},
\end{align}where $\Gamma(x) \equiv \int \nolimits_0^\infty \eta^{x-1}e^{-\eta}d\eta$ is the (complete) gamma function. While the kurtosis for the dye cases is around 3 as expected for a Gaussian distribution, for the ouzo cases it starts with a very high value below $z=$ \SI{3}{\cm}, which corresponds to the initial shear layer in Fig.\ \ref{img:cmap} with $\widetilde{C}_{\text{oil,oversat}} > 1$, and thus we remain doubtful about claiming a regime with super-Gaussian kurtosis. Then the kurtosis sharply decreases to a value close to 2, entering a regime with sub-Gaussian kurtosis, and only slightly increases and approaches 3 as the flow develops. 

The spatial evolutions of the jet width and of the kurtosis in the far field both reveal the radially spread nucleation front following the TNTI in the ouzo jets, whose effect becomes weaker downstream with the consumption of the dissolved oil. Compared with the reference case, the profile with sub-Gaussian kurtosis suggests a wider radial distribution, leading to a more uniform concentration distribution in the domain. Such distribution is in line with the finding of reactive mixing study reported in \citet{Guilbert2021a}, where the reaction makes the gamma probability distribution of the product narrower, expediting the mixing process toward uniformity.

\subsection{Centerline scaling}
\begin{figure}
\centering
\begin{subfigure}[hbt]{0.5\textwidth}
\caption{}
\hspace{-2mm}
\includegraphics[scale=0.8]{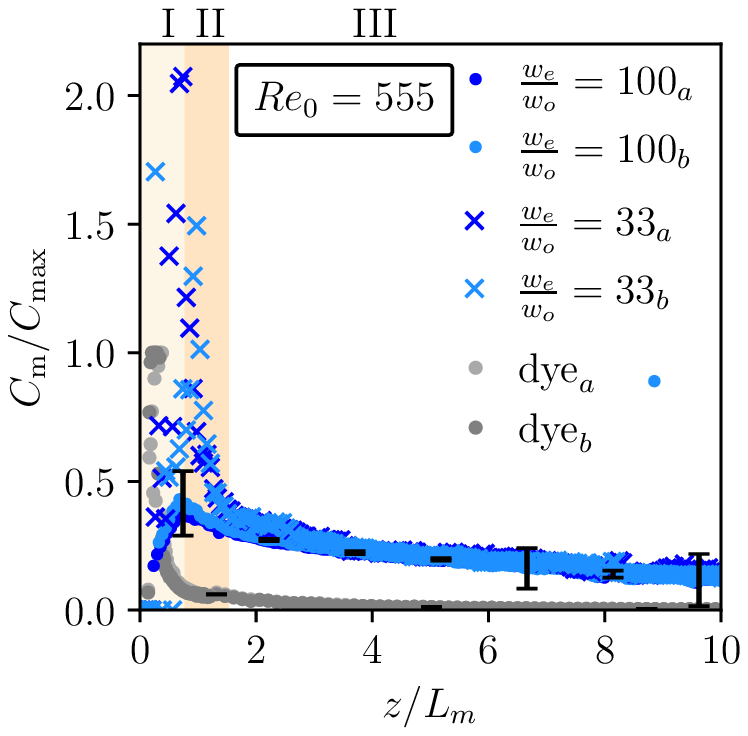}
\label{img:center555}
\end{subfigure}%
\begin{subfigure}[hbt]{0.5\textwidth}
\caption{}
\hspace{-2mm}
\includegraphics[scale=0.8]{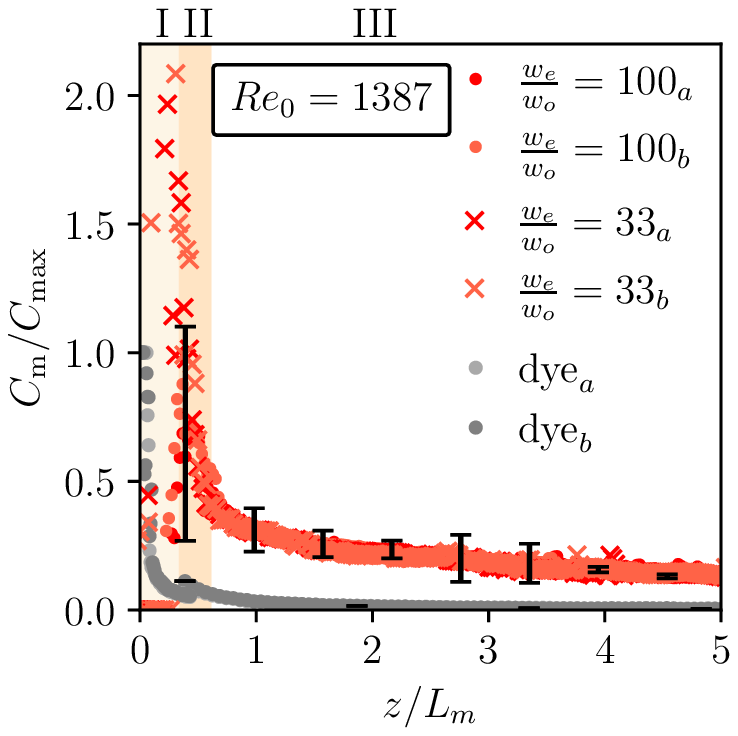}
\label{img:center1387}
\end{subfigure}%
\vspace{-2mm}
\begin{subfigure}[hbt]{0.85\textwidth}
\caption{}
\includegraphics[scale=0.8]{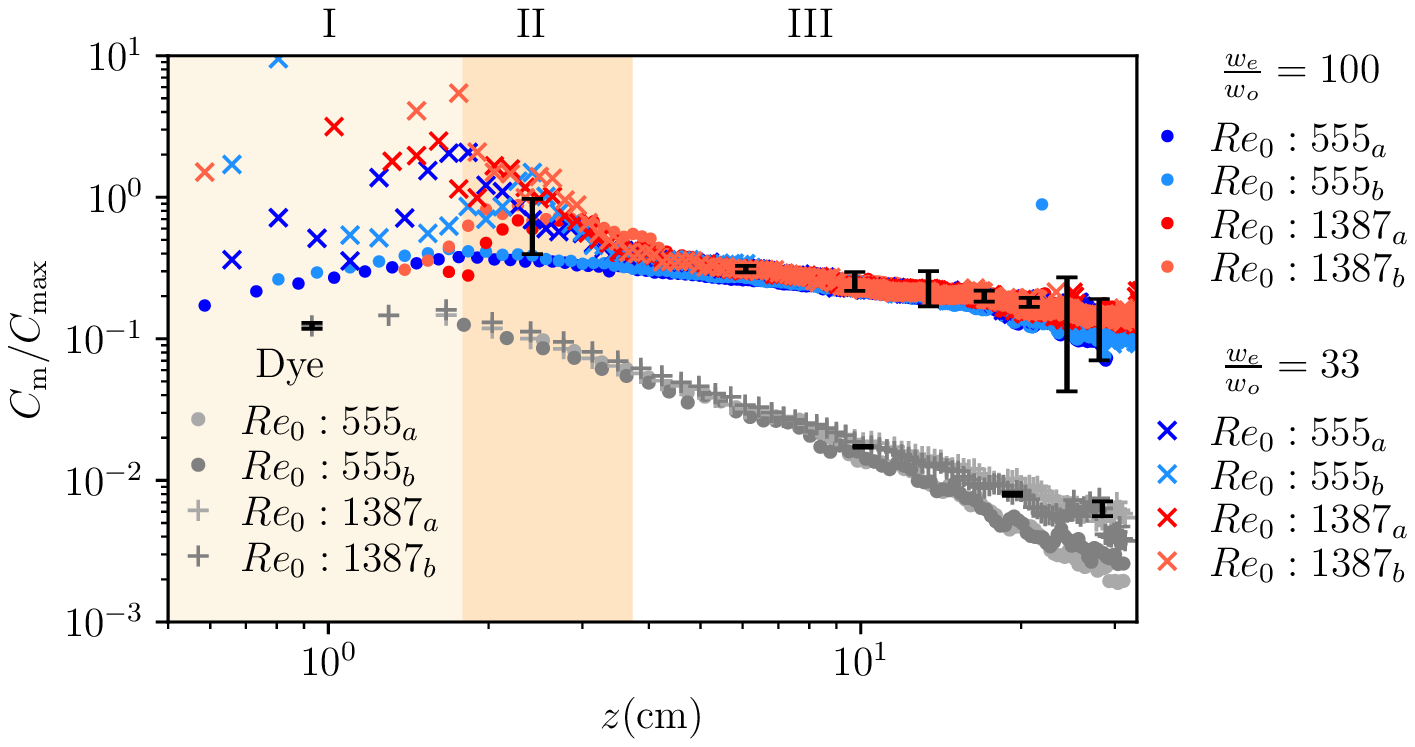}
\label{img:centerlog}
\end{subfigure}%
\caption{Normalized centerline evolution of the rescaled oversaturation for dyed ethanol, ouzo $w_e/w_o=100$, and ouzo $w_e/w_o=33$. (a) and (b) shows results for $Re_0=555$ and $Re_0=1387$, respectively, with normalized height as abscissa. (c) contains all the results in one plot with original height as abscissa. $C_{\text{max}}$ is the initial dye concentration for dye case and the theoretical peak for ouzo cases. The errorbars represent the uncertainty of the curve fitting discussed in \S4.1.}
\label{img:center_total}
\end{figure}

The centerline evolution of the concentration has always been the focus of turbulent jet and plume studies, as the scaling between the dilution and streamwise length is crucial to the self-similarity of the flow. In Fig. \ref{img:center_total} we present the centerline evolution for the dyed ethanol jet and the ouzo jet for different $Re_0$. As the dye concentration dropped monotonically due to dilution, the oversaturation of the ouzo jets experienced several stages, from increasing sharply right above the needle, tightly followed by a sudden decrease, and finally entering a mild dilution regime. In Figs. \ref{img:center555} and \ref{img:center1387} we normalized the height $z$ with the characteristic length scale for the jet-plume transition, $L_m = M^{3/4}/B^{1/2}$, where $M=Q^2/(\frac{1}{4}\pi d^2)$ is the initial momentum flux, and $B=Qg(\rho_{\text{jet}}-\rho_{\text{amb}})/\rho_{\text{amb}}$ is the initial buoyancy flux, see the caption of Table \ref{tbl:condition}. For $z/L_m < 1$, the flow is characterized as momentum dominated jet, followed by a gradual transition to the buoyancy-dominated plume as the flow reaches $z/L_m = 5$. Figs. \ref{img:center555} and \ref{img:center1387} display the centerline evolution for $Re_0 = 555$ and  $Re_0 = 1387$. For the ouzo cases, we can clearly identify three different regimes: (I) a fast nucleation stage where the concentration climbs up quickly. (II) a fast dilution stage, which is only absent for the $Re_0 =555,$ $w_e/w_o=100$ case, and (III) the mild dilution stage in the downstream location, where the nucleation is competing with the dilution. A higher initial oil composition (smaller $w_e/w_o$) induces more intense nucleation in the near field, contributing to the sharper variation in stage (I) and (II) for both $Re_0$. In stage (III), however, the centerline evolution is nearly independent of $w_e/w_o$.

$Re_0$ determines the momentum-buoyancy competition, and in turn the regime where different stages occurs, that is for $Re_0 = 555$ the jet develops into a pure plume regime quickly ($z>5L_m$), while the initial momentum plays a major role for $Re_0 = 1387$. Despite the difference in the regime where the three stages are located, the evolution of the centerline oversaturation only depends weakly on $Re_0$ if the abscissa is not normalized, see Fig. \ref{img:centerlog}. The figure reveals that the effects of $Re_0$ and $w_e/w_o$ are only visible in stage (I) and (II) , while the data from various conditions nearly collapse in stage (III), as $Re_0$ determines the momentum-buoyancy competition which affects the strength of turbulent entrainment, it seems that the centerline oversaturation cannot fully characterize the turbulent entrainment effect on the nucleation. This is not surprising as we believe that the nucleation front follows the TNTI, located at the rim of the jet. In \S4.4 we will show that the effect of the turbulent entrainment on the nucleation is more prominent in the evolution of the oversaturation flow rate which covers the entire domain.

\begin{figure}
\centering
\begin{subfigure}[t]{0.33\textwidth}
\caption{}
\includegraphics[scale=0.7]{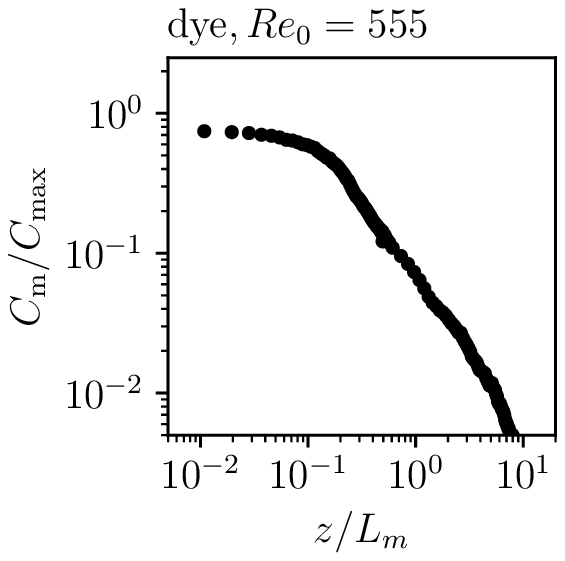}
\label{img:scalingdye_555}
\end{subfigure}%
\begin{subfigure}[t]{0.33\textwidth}
\caption{}
\includegraphics[scale=0.7]{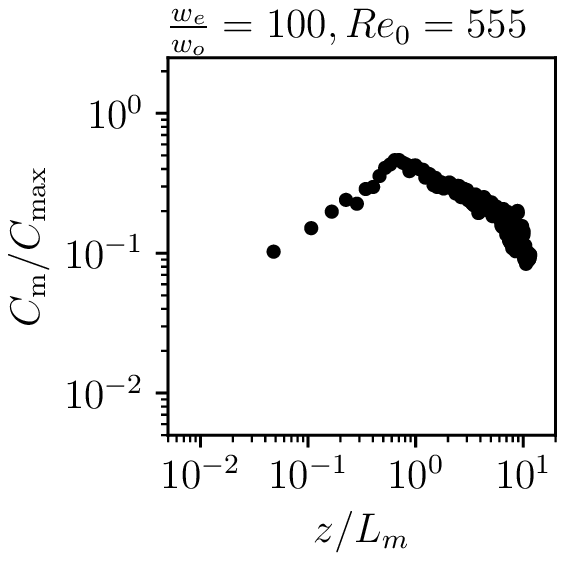}
\label{img:scaling100_555}
\end{subfigure}%
\begin{subfigure}[t]{0.33\textwidth}
\caption{}
\includegraphics[scale=0.7]{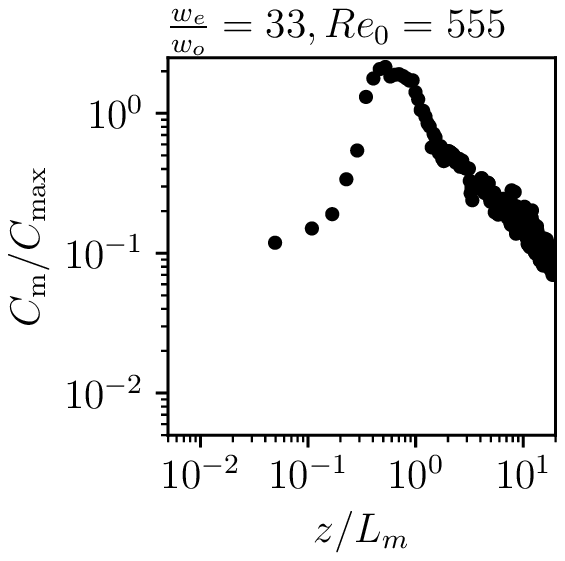}
\label{img:scalin33_555}
\end{subfigure}%
\vspace{-2mm}
\begin{subfigure}[t]{0.33\textwidth}
\caption{}
\includegraphics[scale=0.7]{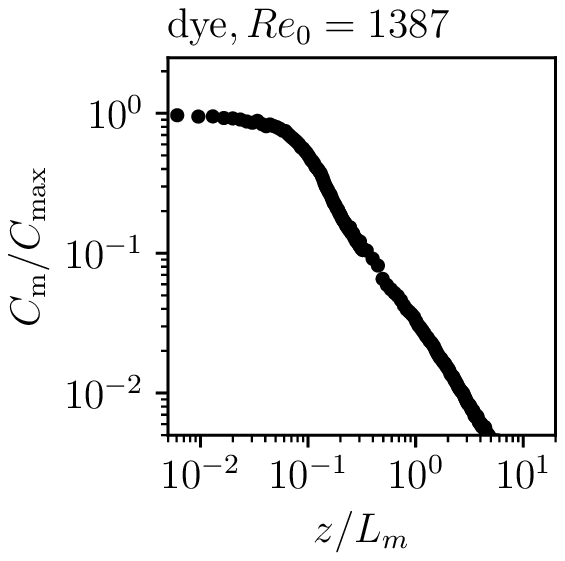}
\label{img:scalingdye_1387}
\end{subfigure}%
\begin{subfigure}[t]{0.33\textwidth}
\caption{}
\includegraphics[scale=0.7]{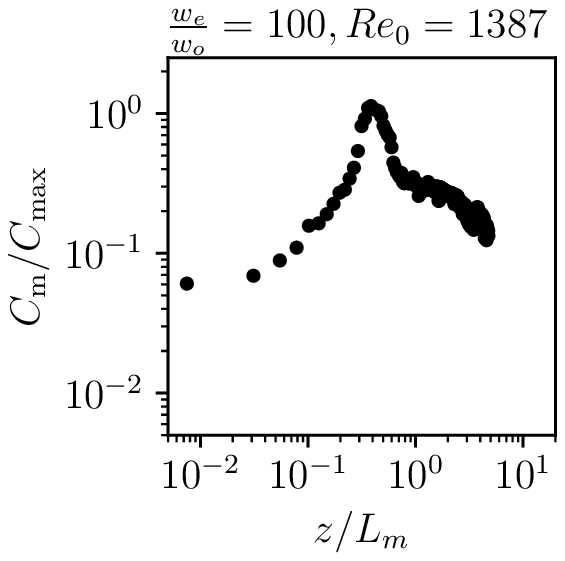}
\label{img:scaling100_1387}
\end{subfigure}%
\begin{subfigure}[t]{0.33\textwidth}
\caption{}
\includegraphics[scale=0.7]{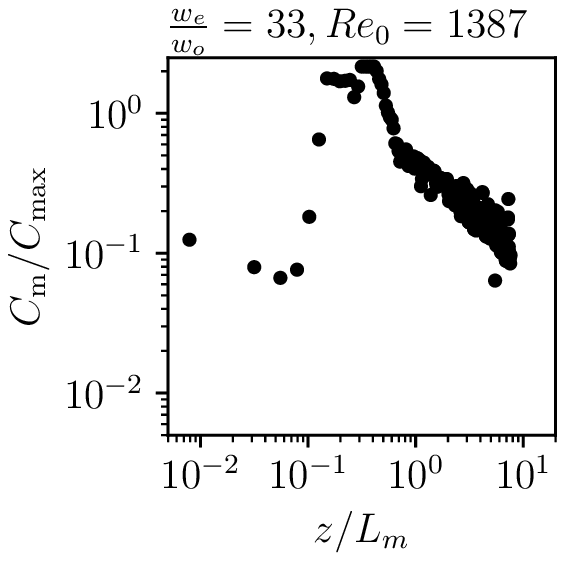}
\label{img:scaling33_1387}
\end{subfigure}%
\vspace{-2mm}
\begin{subfigure}[t]{0.47\textwidth}
\caption{}
\includegraphics[scale=0.8]{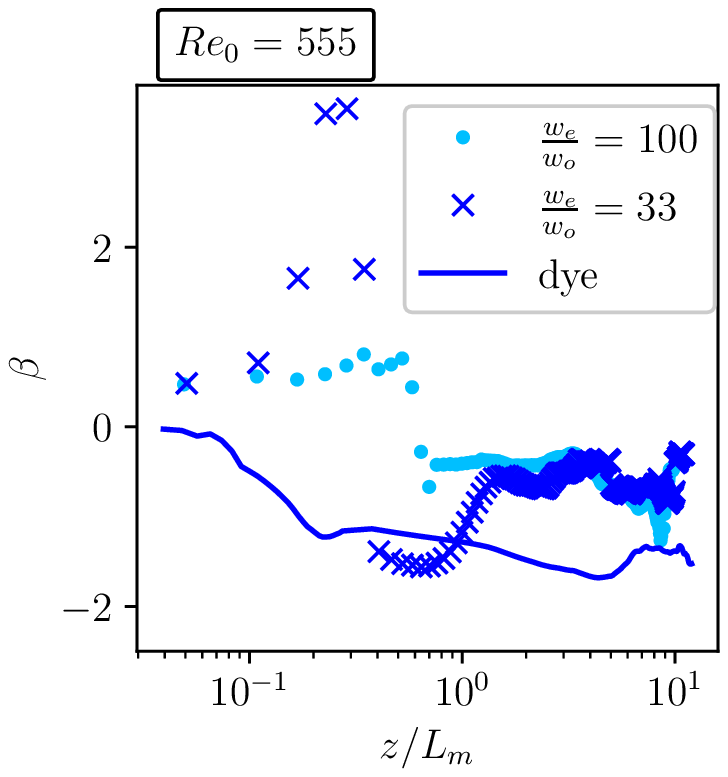}
\label{img:scaling555}
\end{subfigure}\hspace{0.05\textwidth}
\begin{subfigure}[t]{0.47\textwidth}
\caption{}
\includegraphics[scale=0.8]{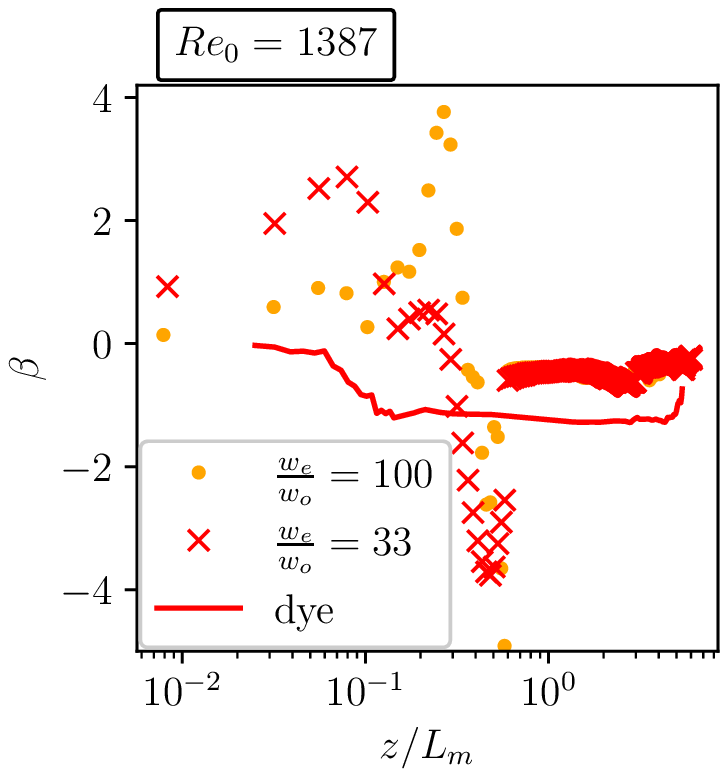}\hspace{0.06\textwidth}
\label{img:scaling1387}
\end{subfigure}%

\caption{Centerline evolution of the oversaturation. Panels (a)--(f) present the centerline evolution in all cases in Fig. \ref{img:center_total}. The curve in each plot is then locally fitted with moving window to deliver the best fitting results. The length of the fitting window is defined as $h_{\text{fit}}$, which starts with $h_{\text{fit}} = 0.1$$z/L_m$ in the fast nucleation regime, changing to $h_{\text{fit}}=$ $z/L_m$ between the peak and $z = 5L_m$, then ending with $h_{\text{fit}} = 3$$z/L_m$ for $z > 5L_m$. The results of the fitting are shown in (g) and (h). $\beta$ denotes the local scaling exponent of $C_m/C_{\text{max}} \propto z^{\beta}$. The ouzo cases can be segmented into three stages as detailed in the caption of Fig. \ref{img:center_total}. In the dilution stage, the exponent $\beta$ ($\beta < 0$ there, representing decaying) is identical for both $w_e/w_o$, and is significantly smaller (i.e., less decaying) than the reference dye case. The fluctuation in the end (large $z$) were probably caused by insufficient averaged frames downstream as described in \S2.3.}
\label{img:scaling_total}
\end{figure}

To retrieve quantitative information from the centerline evolution, it is crucial to know how the oversaturation scales with the height $z$. Therefore the curves with different conditions in Fig. \ref{img:center_total} are separated, while repeated experiments with the same control parameters are ensemble-averaged, leading to Figs. \ref{img:scalingdye_555} through \ref{img:scaling33_1387}, which are then fitted locally using a moving window to derive the local scaling exponent $\beta$ as $C_m(z)/C_{\text{max}} \propto (z/L_m)^{\beta}$. Figs. \ref{img:scaling555} and \ref{img:scaling1387} show that the dye cases match the value in the literature, starting with $\beta = -1$ for the pure jet and then gradually approached the pure plume value $\beta = -1.66$. For the ouzo jets, the multiple stages mentioned above are clearly visible, with $\beta > 0$ in stage (I) initial production stage, then stage (II) a sharp drop to $\beta<-2 $ in a short range, in the end stage (III) reaching a mild dilution regime with $\beta$ slightly below $0$. These transitions reveal the competition between the nucleation and the dilution. Focusing on stage (III), the smaller magnitude of $\beta$ for ouzo jets clearly demonstrates the ongoing and prolonged nucleation even at downstream location. Such finding is similar to the studies in aerosol formation in a turbulent jet by \citet{Lesniewski1998}. They attributed the less steep centerline decay to the nucleation beyond the initial shear layer. We note that the drastic change and the extreme values in the stage (I) and (II) might not be quantitatively accurate, as they are located in the intense nucleation region shown in Fig. \ref{img:cmap}, indicating the violation from the straight diffusion path assumption.

\subsection{Oversaturation flow rate}
\begin{figure}
\centering

\begin{subfigure}[hbt]{0.8\textwidth}
\centering
\caption{}
\includegraphics[scale=0.8]{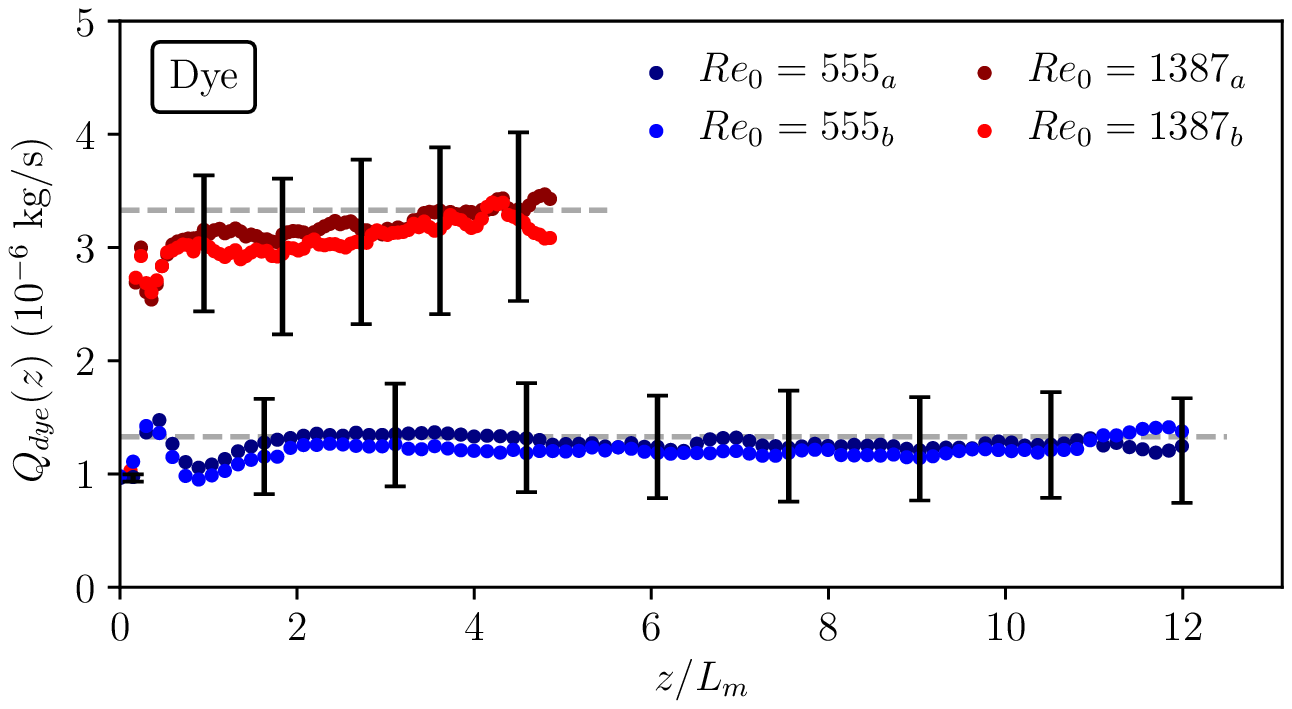}
\label{img:fluxdye}
\end{subfigure}%
\vspace{-5mm}
\begin{subfigure}[hbt]{0.84\textwidth}
\caption{}
\includegraphics[scale=0.8]{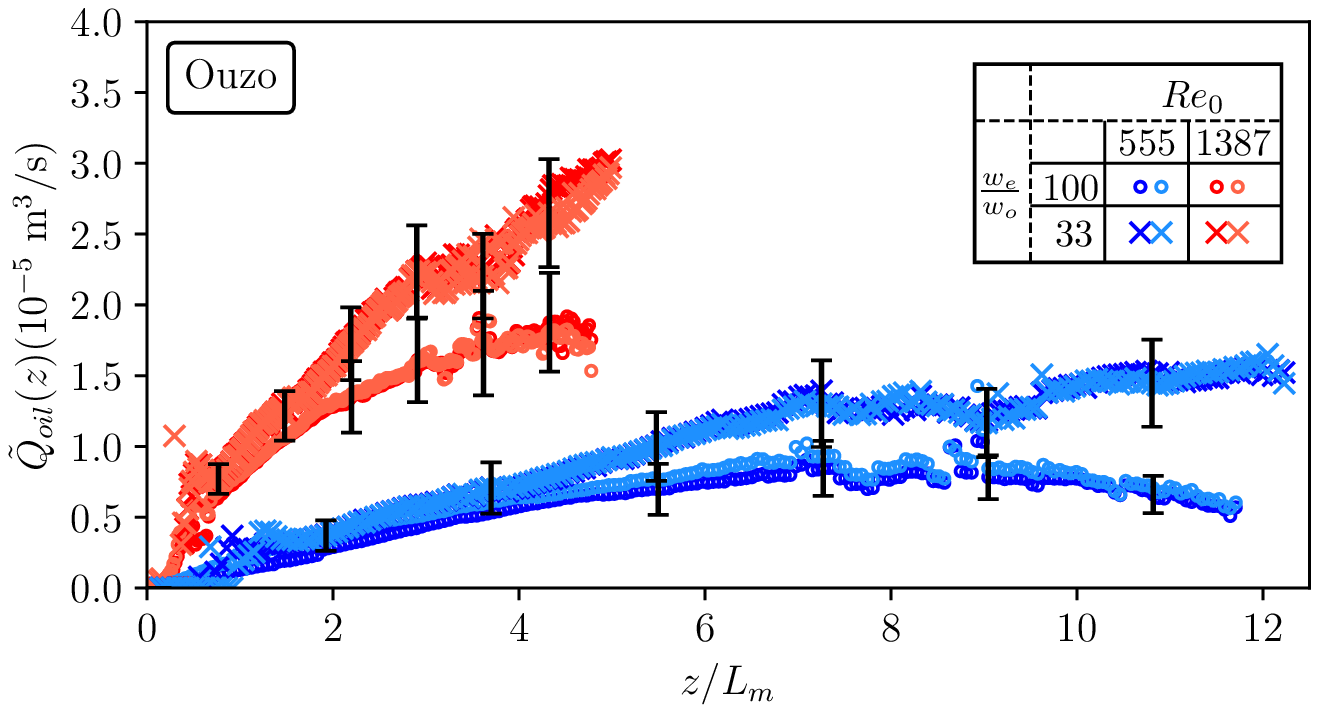}
\label{img:fluxouzo}
\end{subfigure}%
\caption{Mass flow rate evolution along the height. The errorbars show the propagated uncertainty originating from the concentration field. (a) shows the mass flow rate $Q_{dye}(z)$ of reference dye case calculated using Eq. (\ref{eq:fluxdye}), which is reasonably conserved within our domain. The gray dashed lines are the expected values with the initial dye concentration and the volume flow rates. (b) displays the rescaled oversaturation flow rate for the ouzo cases with two different $w_e/w_o$ and two different $Re_0$, calculated using Eq. (\ref{eq:fluxouzo}). The height $z$ is normalized by $L_m$ to show the variation with jet-plume transition. Each legend represents the case with 2 repeating experiments. The results are reproducible while the dependence on $Re_0$ is significant. For $z/L_m \leq 5$ we see the curves between the two compositions are very close, while the composition makes a difference downstream, especially for $Re_0=555$.}
\label{img:flux}
\end{figure}
To make use of the oversaturation data in the entire domain (see Fig. \ref{img:cmap}), we calculated the rescaled oversaturation flow rate by multiplying the oversaturation field, the velocity field from Fig. \ref{img:piv}, and the corresponding area within the discretized annulus region, namely the rescaled oversaturation flow rate
\begin{align}
    \tilde{Q}_{oil}(z) = \sum_{r=0}^{(n-1)dr}\tilde{C}_{\text{oil,sat}}(r,z) u_z(r,z) A_{\text{annu}}(r)
\label{eq:fluxouzo}
\end{align}
Similarly, we can obtain the mass flow rate for the dye reference cases,
\begin{align}
    Q_{dye}(z) = \sum_{r=0}^{(n-1)dr}C_{\text{dye}}(r,z) u_z(r,z) A_{\text{annu}}(r)
\label{eq:fluxdye}
\end{align}
Here $\tilde{C}_{\text{oil,sat}}$ is the rescaled oversaturation from Fig. \ref{img:cmap} (dimensionless) for the ouzo cases and $C_{\text{dye}}$ is the concentration field for dye cases, $u_z(z)$ the streamwise velocity measured by PIV as in Fig. \ref{img:piv}, and $A_{\text{annu}}(r)$ the area of the axisymmetric elements as in Fig. \ref{img:discre}.

For the standard scenario without any nucleation or depletion, the mass flow rate should be conserved in the streamwise ($z$) direction, $Q(z) = const$, which is confirmed in our reference dyed ethanol cases in Fig. \ref{img:fluxdye}. The mass flow rates also match well with the expected values calculated by initial dye concentration (4000 ppm) and the volume flow rates, which are shown by the gray dashed lines. Fig. \ref{img:fluxouzo}, on the other hand, shows the continuous increase of the oversaturation flow rate before reaching the peak. This very nicely demonstrates and quantifies the nucleation. The repeatability of the results across different trials is good, firmly indicating that the nucleation from the solvent exchange process outperformed the dilution within the domain, as both effects resulted from turbulent entrainment. Such finding contrasts the behavior of the centerline scaling exponent $\beta(z)$ in Fig. \ref{img:scaling_total}, as $\beta(z)$ did not become larger than $0$ downstream. The increasing oversaturation flow rate evolution is consistent with the findings in the previous sections, namely the wider radial oversaturation profile and the reduced dilution downstream. Also, the significant dependence of $\tilde{Q}_{oil}(z)$ on $Re_0$ demonstrates the crucial role of the entrainment in the solvent exchange process in the jet, which is not evident in the centerline evolution. We can therefore infer that the oversaturation flow rate characterizes the mean concentration field better than the centerline evolution for this study.

For $z/L_m \leq 2$, that is, before the buoyancy dominates over the momentum,  the experimental data for the two employed compositions $w_e/w_o$ for both $Re_0$ almost collapse, suggesting a potentially universal evolution for the solvent exchange process in this regime. Despite the collapse in the jet regime, the $w_e/w_o = 100$ cases deviate from the $w_e/w_o = 33$ ones with increasing buoyancy dominance, reaching a peak, and even start to decline somewhere downstream for $Re_0=555$. The peak signals the end of nucleation, and the later decline of the curve reflects that the dissolution takes over from nucleation. As discussed in \S2.2 and in Fig. \ref{img:oversat}, when sufficient amount of water is entrained and mixed with the jet fluid, the local water fraction exceeds the critical value corresponding to the upper bound of oversaturation, leading to the dissolution of the nucleated droplets back into the ternary liquid system.
\section{Discussion}
\begin{figure}
\centering

\begin{subfigure}[hbt]{0.5\textwidth}
\caption{}
\includegraphics[scale=0.8]{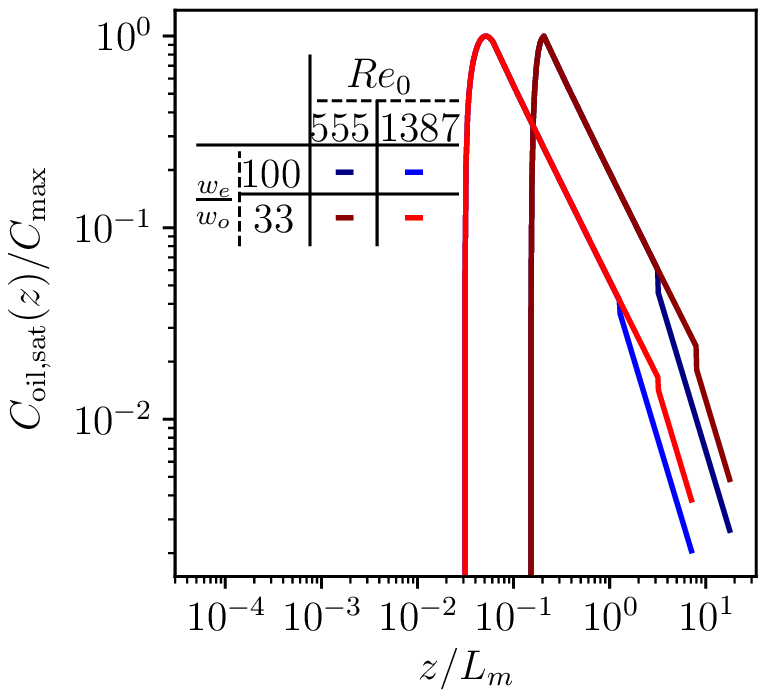}
\label{img:entrain_concen}
\end{subfigure}%
\begin{subfigure}[hbt]{0.5\textwidth}
\caption{}
\includegraphics[scale=0.8]{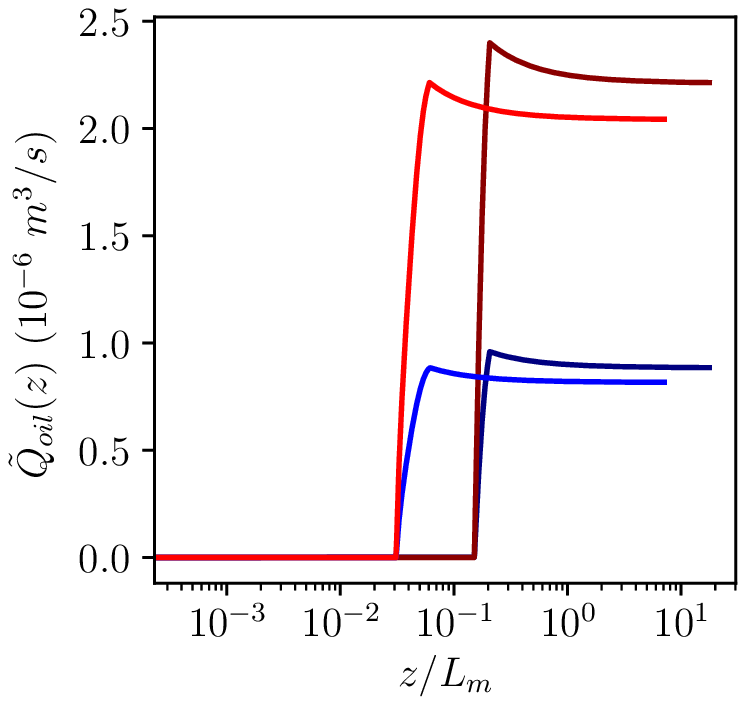}
\label{img:entrain_flux}
\end{subfigure}%

\caption{Analytical results obtained using the entrainment-only model. (a) is determined by the scale analysis, while (b) is obtained by multiplying results of (a) with the volume flow rate, and thus sharing the same plot legend. (a) can be a reference case to the centerline evolution in Fig. \ref{img:center_total}. If the entrained water mixed fully with the jet fluid, the nucleation would complete not far from the needle, which is not observed in the experimental findings. Note that the sharp drop in the figure results from the direct bridging of the volume flow rates for a jet and a plume as mentioned earlier, see Eq. (\ref{eq:qjp}). On the other hand, (b) can be compared with Fig. \ref{img:fluxouzo}, showing the plateau instead of the monotonic increase using the entrainment-only model. } 
\label{img:entrain_flux}
\end{figure}

With the experimental results above, we aim to build up a theoretical framework to capture the observed features, or at least to enable a qualitative explanation of the findings. In the buoyant jet, the turbulent entrainment of water into ethanol-oil mixture is a crucial mechanism for the solvent exchange process, as the nucleation will be activated once the entrained water starts to mix with the injected fluid, forming the ternary system consisting of water, ethanol, and trans-anethole. Incorporating the established turbulent entrainment model, the $z$-direction evolution of the compositions of the elements in the ternary system can easily be obtained. The volume flow rate for a jet ($i=j$) or a plume ($i=p$) is obtained by
\begin{align}
    q_{i}(z) = \pi u_{m,i}(z) (b_{w,i}(z))^2,
\end{align}
where $b_{w,i}=0.115z$ is the jet width for the velocity field from the PIV measurements, and $u_{m,j}(z) = 4.2M^{1/2}z^{-1}$ for the jet, and $u_{m,p}(z) = 3.2B^{1/3}z^{-1/3}$ for the plume from Eq. (\ref{eq:jp}). This results in the volume flow rates 
\begin{align}
    & q_j(z) = 0.17 M^{1/2} z, \nonumber \\
    & q_p(z) = 0.13 B^{1/3} z^{5/3},
\label{eq:qjp}
\end{align}
for the jet and plume, respectively. Here $M$ and $B$ are the initial momentum flux and buoyancy flux, respectively. Note that we directly combined the two models in the middle of the transitional range, $z/L_m = 3$. This procedure is far from perfect but sufficient for a qualitative investigation. With the volume flow rate of the entrained water $q_i(z)$, $i=j,p$, and the volume flow rate of the injected ethanol-oil mixture $q_0$, the oversaturation $C_{oil,sat}(z)$ can be calculated as 
\begin{align}
    & C_w(z) = (q_i(z) - q_0) / (q_i(z) - q_0 + \rho_e q_0), \nonumber \\
    & C_o(z) = \frac{1-C_w(z)}{1+w_e/w_o}, \nonumber\\
    & C_{\text{oil,sat}}(z) = C_o(z) - C_{\text{o,binodal}}(z),
\end{align}
where $C_w(z)$ is the weight fraction of water in the three phase mixture during the process, $C_o(z)$ is the weight fraction of oil in the mixture, and $C_{\text{o,binodal}}(z)$ denotes the saturation concentration of the oil from the binodal curve shown in Fig. \ref{img:binodal}. Fig. \ref{img:entrain_concen} shows the spatial evolution of the oil oversaturation. Although the sharp variation upstream also appeared in the experimental results in Fig. \ref{img:center_total}, the variation here is more pronounced. Also, the oversaturation shows a continuous sharp decrease throughout the entire domain, reflecting that the so-called mild dilution stage in the experimental data is absent in our simplified model. Multiplying the results in Fig. \ref{img:entrain_concen} with the corresponding volume flow rate, as the procedure in Eq. (\ref{eq:fluxouzo}), the overall oversaturation flow rate as in Fig. \ref{img:fluxouzo} can be examined from a theoretical perspective, which is shown in Fig. \ref{img:entrain_flux}. Unlike the (nearly) monotonic increase in Fig. \ref{img:fluxouzo}, the modelled cases here exhibit extremely intense nucleation right above the needle, and the oversaturation flow rate is nearly conserved for almost the entire domain, which indicates the absence of either nucleation or dissolution. The modelled results are similar to findings of \citet{Lesniewski1998} regarding aerosol formation. They identified that the mass flow rate is conserved after the initial shear layer if the nucleation is confined in that area, which is obviously not the case in our study, where we find that nucleation continues to persist throughout the flow.

As the entrainment is the only mechanism we include in the model, the significant deviation from the experimental results unveils that additional factors must contribute to the observed phenomena. \citet{Mingotti2019b} considered mixing as the bottleneck for the fast reactive plume, which led to the prolonged depletion of the injected liquid before reaching stoichiometric level. \citet{Mingotti2019b} pointed out that mixing was constrained within the Batchelor scale located at the rim of the eddies in the reactive plume, causing the incomplete reaction for those entrained fluid. Despite the simplicity of our model lacking the mixing effect for the solvent exchange process, we argue that the weak dilution stage in Fig. \ref{img:center_total} and the monotonic increase in Fig. \ref{img:fluxouzo} are direct consequences of the mixing limitation, considering that the time scale for the fast reaction in \citet{Mingotti2019b} and the micro-droplet nucleation in our study are both extremely short (\citet{Tan2019}). We note that the mixing-limited phenomena not only appeared in a fast reactive plume, but also in other types of multiphase flow, for example the evaporation of droplets in dense sprays (\citet{Rivas2016}, \citet{Villermaux2017}) and in a respiratory puff (\citet{Chong2021}), where the lifetime of the droplets were controlled by the mixing in the ambient humidity field. In our experiments, the limited mixing prevented the oil in the injected fluid from completely nucleating in a short range above the virtual origin, allowing the dissolved oil to be widely distributed by the eddies before nucleation, leading to the wide-spread nucleation across the jet, which has been identified and emphasized throughout our experimental results. 
\section{Conclusions and Outlook}
We have experimentally studied the solvent exchange process in a turbulent buoyant jet. We calculate the oversaturation field from the recorded images using a light attenuation technique, which consists of titration, calibration, axisymmetric discretization, and an optimization algorithm. These experimental procedures involve the background knowledge in the solvent exchange process and the turbulent buoyant jet. The fundamental assumption we made is the straight diffusion path, which leads to the evaluation of the oversaturation, and in turn the nonlinear calibration curve. Except for the region right above the needle, the constructed oversaturation fields gave values below the theoretical upper bound for the oversaturation, consistent with the diffusion path assumption. In addition, the very similar results between the two cameras further support the validity of the assumption, the implemented methods, and hence the calculated oversaturation fields. 

Analyzing the rescaled oversaturation fields, the ouzo jets exhibited wider jet spreading $b_T(z)/z$ and a radial profile with sub-Gaussian kurtosis, reflecting enhanced nucleation in the entrainment region. In the streamwise direction, the centerline oversaturation demonstrated a three-stage evolution, consisting of fast nucleation, sharp dilution, and finally mild dilution, with the scaling exponent $\beta$ of the centerline oversaturation being almost independent of $Re_0$ and the initial compositions of the ethanol-oil mixture, $w_e/w_o$, in the final stage. In contrast, integrating the velocity field, the oversaturation flow rates in the entire domain show monotonic increase before reaching the peak, again reflecting the wide-spread nucleation of the oil droplets across the jet in the entrainment region.

In addition to the experimental findings, we have formulated a simplified model, only considering the turbulent entrainment. It serves as the first step to build the theoretical framework for the ouzo jet. The differences between the experimental results and the model suggest the existence of other mechanisms, which we attributed mainly to the rate-limiting mixing constraint within the Batchelor scale. The entrainment rate exceeded the mixing rate, leading to the prolonged nucleation downstream and across the jet.  

To the authors' knowledge, this work is the first attempt to experimentally and theoretically tackle the ouzo jet, a complex fluid mechanics problem combining solvent exchange and turbulent shear flow. The approach to quantify the concentration of the nucleated droplets, namely the oversaturation, seems a promising way for future efforts in this line of research. However, we are aware of the limitation of the proposed method, which is a lack of direct measurements for droplets size distributions and for the local concentration. To the authors' knowledge, estimation of nucleated droplet size and number distribution in the early studies is limited to small scale system, allowing the use of microscopic instruments, and thus free from the difficulties such as struggling between depth of field and resolution. With the limitations in large scale and turbulent flow, we have no choice but to focus on temporally averaged and depth integrated images first before diving into local and fluctuating micro-scale characteristics.

Attempting to see through the opaque flow induced by solvent exchange, we plan to conduct future research in a laterally confined quasi two-dimensional jet, hoping to reveal the time-dependent process and the local characteristics missing in the current study. Also, as TNTI is mentioned several times, we hope to visualize the phenomena by reducing the opaqueness of the flow in the quasi two-dimensional jet \citep{Giger1991,Watanabe2015b,Rocco2015}. Last but not least, an analytical framework including the mixing rate would be highly valuable towards a thorough understanding of the topic.
\vspace{\baselineskip}

\section*{Acknowledgement}
The authors acknowledge the funding by ERC Advanced Grant Diffusive Droplet Dynamics with Project No. 740479, Natural Science Foundation of China under grant Nos. 11988102 and 91852202 and Netherlands Organisation for Scientific Research (NWO) through the Multiscale Catalytic Energy Conversion (MCEC) research center.

We thank G.W. Bruggert, M. Bos, and T. Zijlstra for technical support building the setup. Valuable discussions with R.A. Lopez de la Cruz, L. Thayyil Raju, H. Tan, and Y. Li are also appreciated.

\section*{Declaration of Interests}
The authors report no conflict of interest.
\newpage
\section*{Appendix A: Details of the optimization process}
In the optimization process, the loss function is set to the square of the difference of left-hand side and right-hand side of Eqs. (\ref{eq:LAdye}) and Eq. (\ref{eq:LAouzo}). A basinhopping algorithm is used to search for global minimum. We gradually increase the resolution, namely we decrease the size of the discretization unit $dr$ as in Fig. \ref{img:discre}, and finally obtain the results with $dr=2$ pixels. In this way we get a rough, low-order, low-resolution estimation for the oversaturation, which we then refine by decreasing $dr$, where we upscale the low-resolution oversaturation profile to a higher resolution, which is our new starting point for the minimization. This incremental procedure allows us to find global minima for our high-dimensional system in an efficient manner. The calculation procedure is repeated for every height independently, leading to the oversaturation fields as in Fig. \ref{img:cmap}.

\begin{figure}
\centering
\begin{subfigure}[hbt]{1\textwidth}
\centering
\includegraphics[scale=0.8]{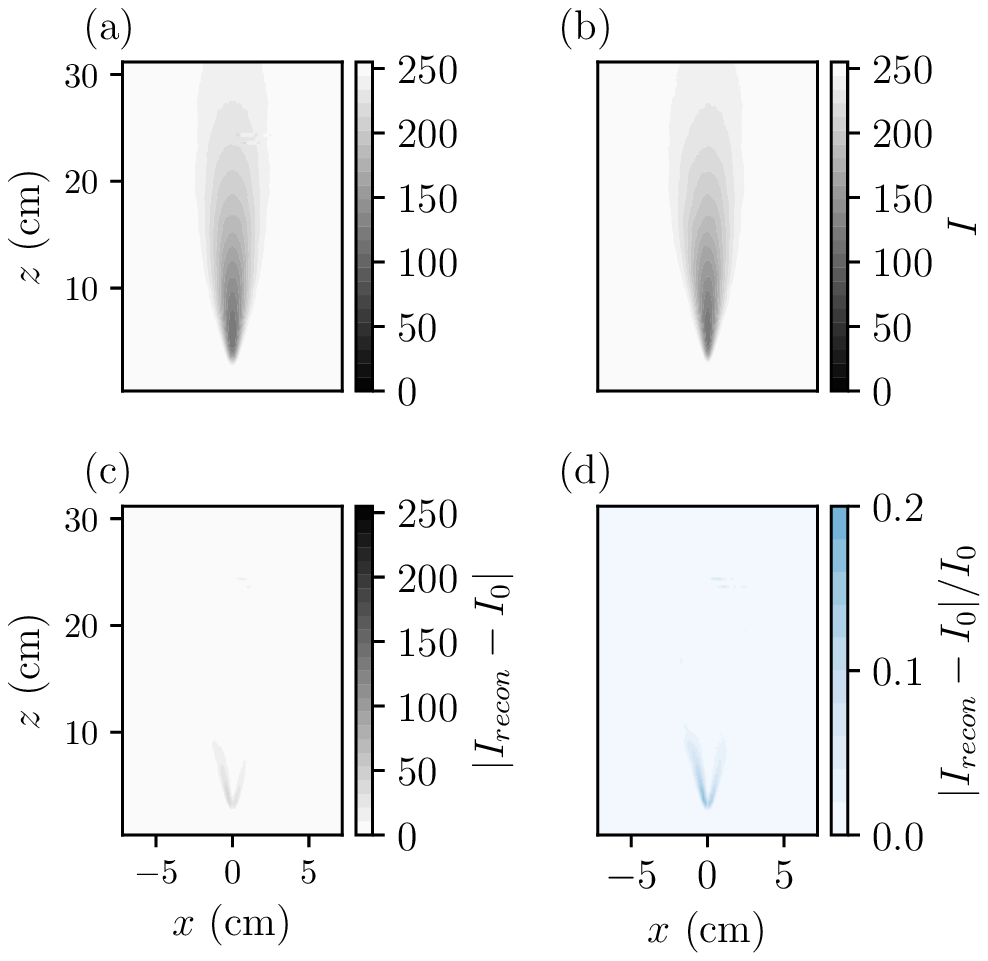}
\label{img:optdq20}
\end{subfigure}%
\vspace{-2mm}
\begin{subfigure}[hbt]{1\textwidth}
\centering
\includegraphics[scale=0.8]{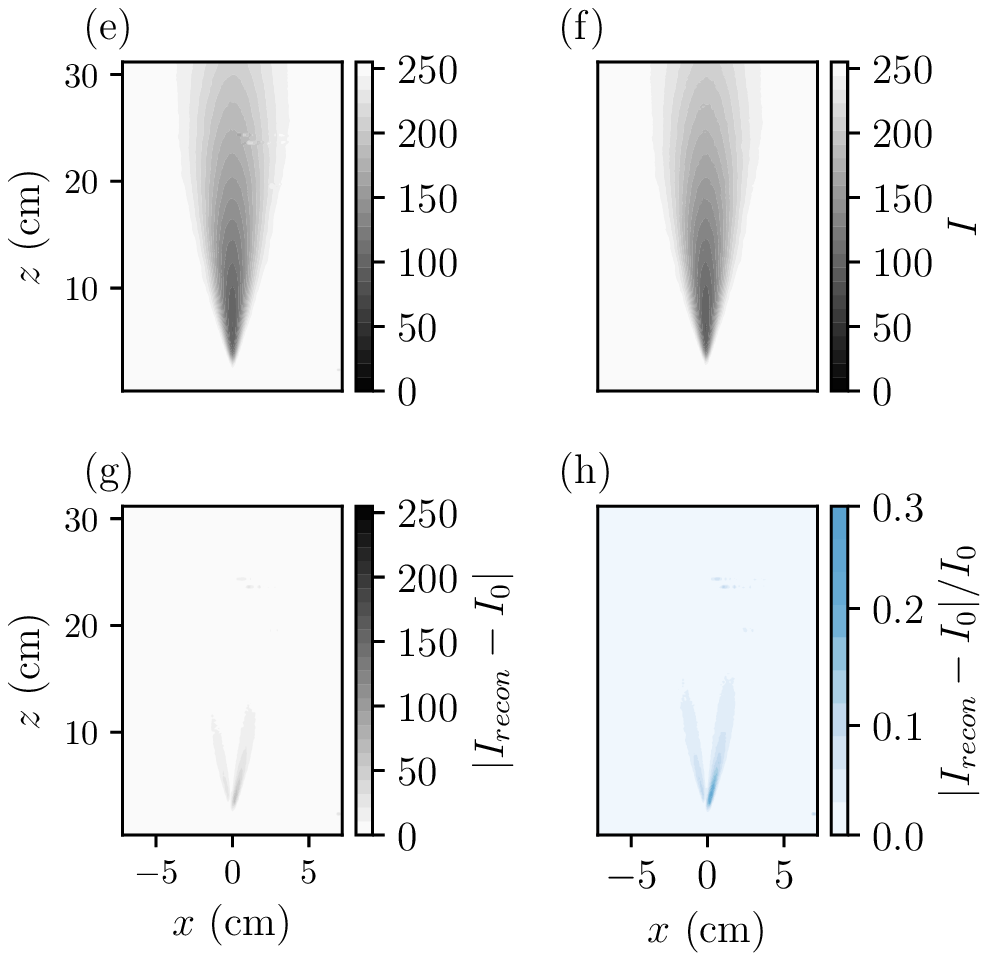}
\label{img:optdq50}
\end{subfigure}%
\caption{Evaluation of the oversaturation fields calculated by optimization procedures. (a-d) $Re_0$=555, $w_e/w_o$=100, (e-h) $Re_0$=1387, $2_e/w_o$=100. (a,e) are the reconstructed intensity fields by the oversaturation fields. (b,f) are the original intensity fields recorded by the camera. (c,g) present the absolute difference between the reconstructed and original intensity fields. (d,h) show the intensity difference in fraction, namely the results in (c,g) divided by those in (b,f). Note that the colorbars for (c,g) are inverted. }
\label{img:optd}
\end{figure}

To check the validity of the results, the final oversaturation field is fed back into Eq. (\ref{eq:LA1}) to reconstruct the intensity field, see Figs. 15a and 15e. The difference between such reconstructed field and the original intensity field (Figs. 15b and 15f) can then be used to check the deviation, see Figs. 15c, 15d, 15g, and 15h. Note that for direct comparison, the original intensity field is down-sampled to match that of the reconstructed field. Fig. \ref{img:optd} clearly shows that the results are not perfect, with most deviation located in the near field. We believe that there are multiple factors contributing to the deviation, mainly including the down-sampling of the original intensity field, the resolution limit in the near field, and the vicinity of the region with $\widetilde{C}_{\text{oil,oversat}} > 1$. Considering the aforementioned factors, and the low deviation in most of the domain, we believe the results obtained by the optimization process are valid for the semi-quantitative analysis in this study.

\section*{Appendix B: Comparison between far and zoomed-in recordings}
\begin{figure}
\hspace{1mm}
\begin{subfigure}[H]{1\textwidth}
\includegraphics[scale=0.8]{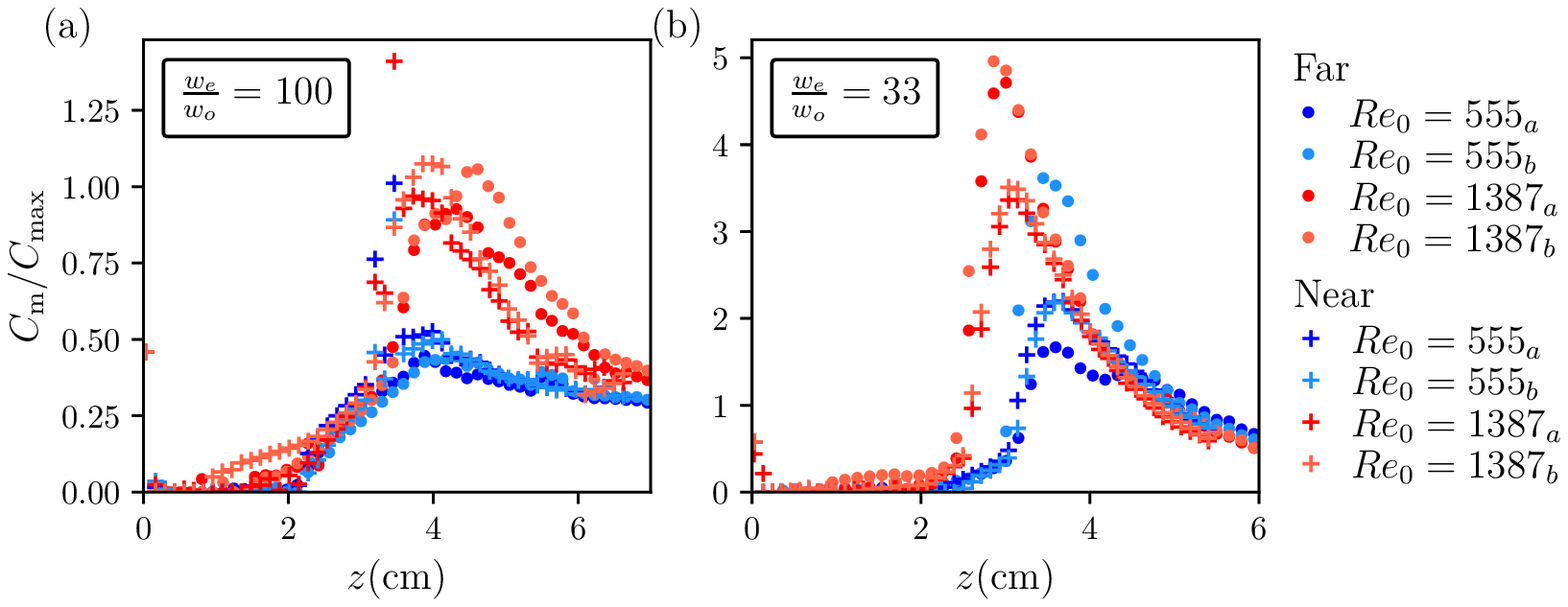}
\label{img:fn_dir}
\end{subfigure}
\begin{subfigure}[H]{1\textwidth}
\hspace{2mm}
\includegraphics[scale=0.8]{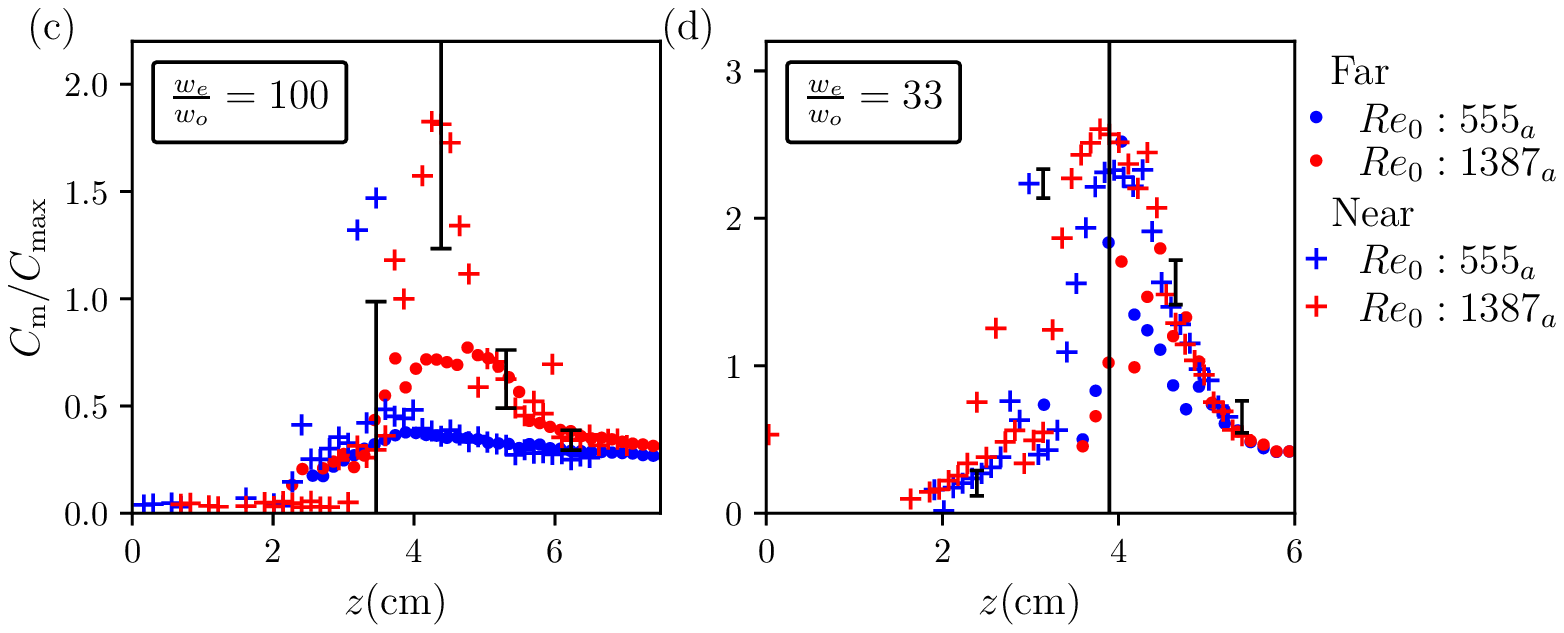}
\label{img:fn_fit}
\end{subfigure}\vspace{-3mm}

\begin{subfigure}[H]{1\textwidth}
\hspace{3mm}
\includegraphics[scale=0.8]{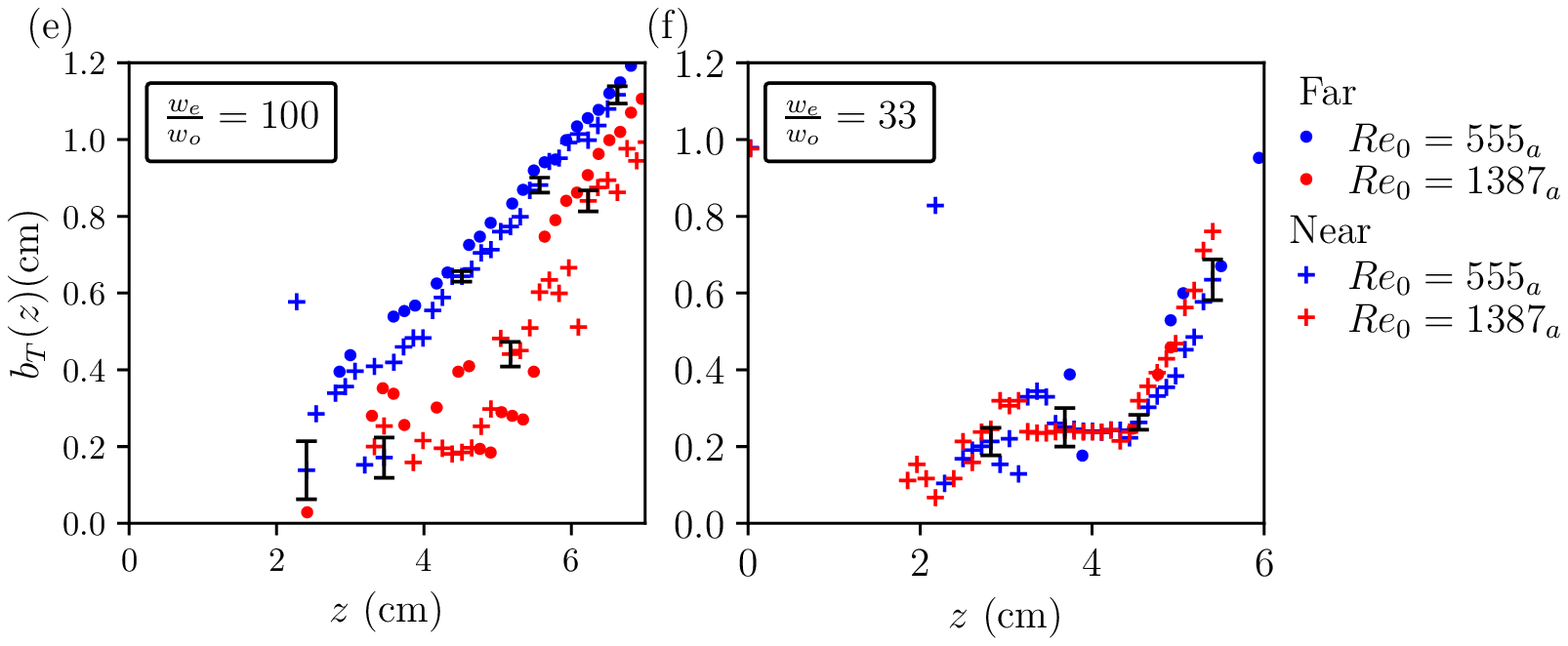}
\label{img:fn_sp}
\end{subfigure}\vspace{-2mm}
\caption{Comparison of oversaturation profiles between far view and zoomed-in recordings. (a,b) show the streamwise profiles, namely the centerline evolution directly derived from the calculated fields in Fig. \ref{img:cmap} , and (c,d) are the centerline evolution obtained by curve-fitting. (e,f) display the radial profile, namely the the jet width evolution, obtained by curve-fitting.}
\label{img:fn}
\end{figure}

Fig. \ref{img:fn} compares the profiles obtained from far view and zoomed-in recordings in streamwise and radial directions. Figs. 16a and 16b presents the centerline evolution of the results in Fig. \ref{img:cmap}, showing nice collapse between the two fields except for the region with $\widetilde{C}_{\text{oil,oversat}} > 1$. We also compare the centerline evolution obtained from the curve-fitting in Figs. 16c and 16d, with similar features to those in Figs. 16a and 16b. Note that for the intense nucleation region, the error bar is extremely large because the data points for fitting are scarce, demonstrating the uncertainty of our results in the area. Also, one might notice the fluctuation for each case in the ramping-up stage, which can be attributed to the fluctuating position of laminar/turbulent transition, causing the nucleation there not so smooth. To complete the comparison, the jet width evolutions are presented in Figs. 16e and 16f, showing consistent radial profiles between the two recordings. Note that we do not subtract the abscissa with laminar length here \citep{Hassanzadeh2021}, unlike the analysis in \S4, so that the height $z$ here is aligned with that in Fig. \ref{img:cmap}. By the rigorous comparison between the two recordings, the consistency of our method is further demonstrated for the domain out of the initial shear layer.

\section*{Appendix C: Light attenuation in a scattering medium}
As briefly mentioned in \S.2.3, the superposition of level of light attenuation is questionable in a scattering medium. \citet{Dahm2013a,Dahm2013b} discuss the way to theoretically address the light propagation in such medium. The light attenuation technique we implemented is based on the Lambert-Beer law, see eq. (\ref{eq:LA0}). For a non-scattering medium, the fraction of light propagating through a medium is called transmittance, while the remaining fraction of light not going through is absorbance. However, with oil droplets induced by solvent exchange, the solution becomes a scattering medium, which introduces an extra factor called remittance in \citet{Dahm2013a,Dahm2013b}, that is,
\begin{equation}
    T+A+R = 1,
\label{eq:TAR}
\end{equation}
where $T = I/I_{ref}$ is transmittance, A is absorbance, and R is remittance.  The remittance represents the part of light that neither penetrates through the medium nor absorbed by the medium, which results from its scattering nature. Instead of using a linear superposition, \citet{Dahm2013b} introduces the so-called equations of Benford to deal with the overall transmittance, absorbance, and remittance through layers of the scattering medium. These equations are written as:

\begin{align}
    & T_{x+y} = \frac{T_x T_y}{1-R_x T_y}, \nonumber \\
    & R_{x+y} = R_x + \frac{T^2_x R_y}{1-R_x R_y}, \nonumber\\
    & A_{x+y} = 1 - T_{x+y} - R_{x+y},
\end{align}
where x, y denote the two adjacent layers of the scattering medium with their own thickness and concentration.

While the remittance R is not possible to measure in our setting, we measure the transmittance T as a function of oil oversaturation with two calibration cells of difference thickness (10mm and 2mm), namely $T_x(C)$ and $T_y(C)$. And from $T_x(C)$ and $T_y(C)$ we can obtain $R_x(C)$ and $R_y(C)$ from the equations of Benford. At this point we have the transmittance and remittance curves as functions of oil oversaturation at the thickness of 2mm and 10mm. Using either of these two curve sets and the equations of Benford, we can obtain the transmittance and remittance curves T(C) and R(C) for a very thin layer of the medium for further calculation. Here we choose 1 pixel as the thickness for such base layer. Applying the thickness matrix $dy_{mn}$ in eq. \ref{eq:LAouzo}, we can estimate the amount of the base layers in each item of $dy_{mn}$, from which we can construct the overall transmittance $T_{sum}$ and remittance $R_{sum}$ using the equations of Benford. $T_{sum}$ needs to match the value from the averaged jet in Fig. \ref{img:meanimg}. Note that the order/trajectory of light propagation matters when applying the equations of Benford, which is not the case for the Lambert-Beer law. To elaborate the procedure further, we now take a look at Fig. \ref{img:discre}. To estimate $T_{sum}$, we have to start from the discretization unit closest to the light source. Knowing the amount of base layers (1 px) in this unit, the overall transmittance till the end of this unit can be obtained, and then we enter the next unit.  When applying the equations of Benford, the value $T_x$ keeps getting updated with the results obtained from the previous step, while $T_y$ is the value of the following base layer, which could be in the same or the next unit. We repeat such algorithm following the direction of light propagation, reaching $T_{sum}$ right in front of the camera. 

We then implement such theoretical optical framework into the optimization procedures in Appendix A, replacing the superposition described by eq. \ref{eq:LAouzo}. Unfortunately, with the extra complexity, the optimization procedure is unable to converge to a sensible concentration field.

Considering the discussion in this appendix, we must remind the readers that the preliminary analysis and results in this work is not on firm and water-proof footing, and will require further examination once the updated algorithm can successfully capture the optics involved in this complex phenomena.


\bibliographystyle{jfm}
\bibliography{jfm-instructions}

\end{document}